\documentstyle[aps,psfig]{revtex}
\newcommand{\bra}[1]{\mbox{$\langle #1|$}}
\newcommand{\ket}[1]{\mbox{$|#1\rangle$}}

\newcommand{\vpr}[2]{\mbox{$\vec{#1}\!\cdot\!\vec{#2}$}}

\begin{document}
%\preprint{MKPH-T-00-12}

\title{
\hfill{\small {\bf MKPH-T-00-12}}\\
{\bf The Role of Meson Retardation in Deuteron Photodisintegration 
above Pion Threshold}
  \footnote[2]
  {Supported by the Deutsche Forschungsgemeinschaft (SFB 443).}
}
\author
{Michael Schwamb and  Hartmuth Arenh\"ovel}
 \address{ Institut f\"ur Kernphysik,           
  Johannes Gutenberg-Universit\"at,  
  D-55099 Mainz, Germany }
\maketitle

\begin{abstract}
\noindent
Photodisintegration of the deuteron above $\pi$ threshold is studied in a 
coupled channel approach including $N \Delta$  and $\pi d$ channels with 
consideration of pion retardation in potentials and exchange currents. 
A much improved description of total and differential cross sections in 
the energy region between $\pi$ threshold and 400-450 MeV is achieved. 
With respect to polarization observables, the description of the 
linear photon asymmetry and the proton polarization remains problematic.
\end{abstract}

\pacs{PACS numbers: 21.45.+v, 13.40.-f, 25.20.-x}
%13.75.Cs, 21.45.+v, 24.10.Eq, 25.10.+s}

\section{Introduction}\label{kap5_vor}

Recently we have constructed in \cite{ScA99}, henceforth denoted as I, a 
realistic hadronic interaction model in the two-nucleon system which takes 
into account meson retardation completely. This model is based on meson, 
nucleon and $\Delta$ degrees of freedom, and it is able to describe 
$NN$ scattering quite satisfactorily 
also above pion threshold. However, for technical reasons it only allows 
configurations where maximal one meson is present explicitly at a time. 
Within this restriction, full meson retardation in the potentials has been 
incorporated in a consistent manner. The motivation for this work was 
twofold: First of all, from the general principle of relativity any 
interaction has to be retarded taking into account the finite velocity 
of its propagation. With respect to the $NN$ interaction, this is 
particularly evident for energies above pion threshold, where an exchanged 
pion can become an onshell particle, so that the retarded meson propagator 
becomes singular describing the coupling of the $NN$ channel to the now open 
$\pi NN$ channel. Secondly, there is strong evidence in deuteron 
photodisintegration above $\pi$ threshold, in particular in the 
$\Delta$ resonance region, that neglected retardation is one major cause for 
the failure to describe experimental data by quite sophisticated theoretical 
treatments which, however, are based on static interactions and corresponding 
static exchange currents. Indeed, first 
results have demonstrated clearly the considerable improvement of the 
theory if retardation is considered~\cite{ScA98,Sch99}. 
 
In the present work, we have studied deuteron photodisintegration in 
greater detail with respect to the influences of the various ingredients 
of the theoretical framework developed in I. In Sect.~\ref{current} we 
discuss the general structure of the e.m.\ current operator, which consists 
of the baryonic components, for example, the one-body spin, convection, and 
isobar currents, as well as of mechanisms where mesons, especially pions, are 
created or absorbed. The latter ones are a natural consequence of the 
incorporation of mesons as explicit degrees of freedom. As is well known, 
the internal absorption of a photon by a nucleon creating an intermediate 
$\Delta$ isobar is of particular importance in the $\Delta$ resonance region. 
Therefore, Sect.~\ref{kap5_m1plus} is devoted to a careful determination of 
the corresponding $\gamma  N \Delta$ coupling by studying the $M_{1+}^{(3/2)}$ 
multipole in pion photoproduction on the nucleon. In this way, {\it all} 
parameters of the theory are fixed in advance before studying electromagnetic 
reactions in the two-nucleon system. In the following 
Sect.~\ref{kap5_effektiv_strom} the resulting effective current, acting in the 
baryonic space only, is derived. Special emphasis is devoted to the structure 
of the retarded MEC which has, in contrast to its static counterpart, a 
singular structure above $\pi$ threshold describing the coupling of the $NN$ 
to the $\pi NN$ channel. In Sect.~\ref{kap5_eichinvarianz_2} the question of 
gauge invariance is discussed. The results for total and differential cross 
sections as well as for selected polarization observables are presented in 
Sect.~\ref{kap6_deutspalt}. Finally, Sect.~\ref{summary} contains a summary 
and an outlook.

\section{The electromagnetic current operator}\label{current} 

The transition amplitude $T_{fi}$, which describes the absorption of a 
photon by a hadronic system making a transition from an initial state 
$\ket{\vec{p}_i,i}$ to a final state $\ket{\vec{p}_f,f}$, is given in 
Coulomb gauge by the Fourier component of the e.m.\ current matrix element 
${\vec J}_{fi}(\vec{k})$
\begin{equation}\label{tfi}
  \delta^{(3)}(\vec{p}_f - \vec{p}_i- \vec k) T_{fi} = -
 \vec{\epsilon}\,(\vec{k},\lambda) \cdot   {\vec J}_{fi}(\vec{k})\,,
\end{equation}
where the polarization vector of the incoming photon with momentum $\vec k$ 
and helicity $\lambda$ is denoted by $\vec{\epsilon}\,(\vec{k},\lambda)$. 

\subsection{The general structure of the current}\label{struc_current}

\begin{figure}[htp]
\centerline{\psfig{figure=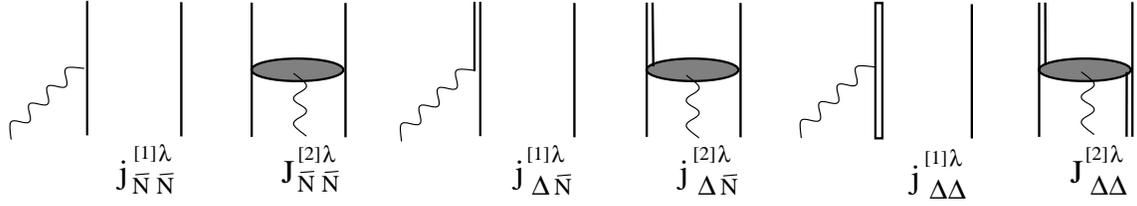,width=15cm,angle=0}}
\vspace{0.5cm}
\caption{Diagrammatic representation  of the baryonic currents. A shaded 
ellipse symbolizes a two-body exchange current.}
\label{figem1}
\end{figure}

We begin with a brief discussion of the formal structure of the e.m.\ current.
As has been described in detail in I, the model Hilbert space of the 
two-nucleon system consists of three orthogonal subspaces
 \begin{equation}\label{kap3_hilbert}
{\cal H}^{[2]} = {\cal H}^{[2]}_{\bar N} \oplus
{\cal H}^{[2]}_{\Delta} \oplus 
{\cal H}^{[2]}_{X}\,
\end{equation}
containing either two bare nucleons $({\cal H}^{[2]}_{\bar N})$,
one nucleon and  one $\Delta$ (${\cal H}^{[2]}_{\Delta})$, 
or two nucleons and one  meson $({\cal H}^{[2]}_{X})$.
Accordingly, the current operator can be decomposed into various diagonal 
and non diagonal components. With the help of the projection operators 
$P_{{\bar N}}$, $P_{\Delta}$, and $P_{X}$ onto the above subspaces, 
respectively (see I), the current $J^{\mu}(\vec{k})$ can be written as a 
symbolic $3\times 3$ matrix
\begin{equation}\label{kap5_strom_matrix}
J^{\mu}(\vec{k})
 = \left( \begin{array}{ccc}
               J^{\mu}_{{\bar N}{\bar N}}(\vec{k})
 & J^{\mu}_{{\bar N}\Delta}(\vec{k}) & J^{\mu}_{{\bar N} X}(\vec{k}) \\
 J^{\mu}_{\Delta {\bar N}}(\vec{k}) & J^{\mu}_{\Delta \Delta}(\vec{k})
 & J^{\mu}_{\Delta X}(\vec{k}) \\
          J^{\mu}_{X {\bar N}}(\vec{k}) & J^{\mu}_{X \Delta}(\vec{k}) &
 J^{\mu}_{X X}(\vec{k}) 
 \end{array} \right)   \,.
\end{equation} 
Now, we will discuss the various components in detail. We distinguish 
between pure ``baryonic'' currents, ``one-meson production/annihilation'' 
currents, 
and the remaining currents, and divide the first one further into one- 
and two-body operators, labeled by a superscript ``[1]'' and ``[2]'', 
respectively. The diagonal baryonic currents 
$J^{\mu}_{{\bar N}{\bar N}}(\vec{k})$ and 
$J^{\mu}_{\Delta \Delta}(\vec{k})$ are represented diagrammatically 
in Fig.~\ref{figem1}. The two-body parts comprise effective exchange 
currents, like for example, heavy meson exchange currents, which are not 
generated explicitly via the elementary meson-baryon 
vertices and the meson production/annhilation currents. The transition current 
$J^{\mu}_{\Delta {\bar N}}(\vec{k})$, also represented in 
Fig.~\ref{figem1}, describes the e.m.\ excitation of a 
$\Delta$ resonance including two-body contributions. 

\begin{figure}[htp]
\centerline{\psfig{figure=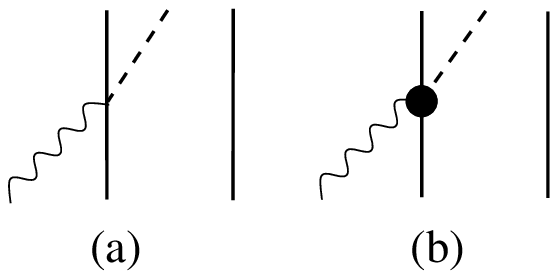,width=6cm,angle=0}}
\vspace{0.5cm}
\caption{Diagrammatic representation of the meson production currents 
$J^{\mu}_{X{\bar N}}$: (a) contact current $j_{X\bar N}^{(1)\,\mu}$ and 
(b) vertex current $j_{X\bar N}^{(1v)\,\mu}$.}
\label{figem2}
\end{figure}

\begin{figure}[htp]
\centerline{\psfig{figure=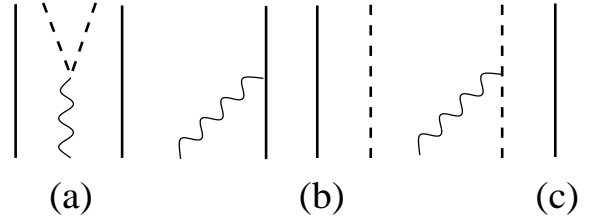,width=8cm,angle=0}}
\vspace{0.5cm}
\caption{Diagrammatic representation of the two-meson production current 
$j^{(0)\, \mu}_{X{\bar N} }$ (a) and the current components 
$J^{\mu}_{XX}$: (b) nucleon current $j^{{\bar N}\, \mu}_{XX}(\vec{k})$,
  (c) meson current   $j^{X \, \mu}_{XX}(\vec{k})$.}
\label{figem3}
\end{figure}

The other transition current $J^{\mu}_{X {\bar N}}$ consists of three 
components, namely 
$j^{(0)\, \mu}_{X {\bar N}}$, $j^{(1)\, \mu}_{X {\bar N}}$, and 
$j^{(1v)\, \mu}_{X {\bar N}}$, where the superscript ``(0)'' or ``(1)'' 
indicates the order with respect to the meson-nucleon coupling constant. 
Of these, the one-meson production currents $j^{(1)\, \mu}_{X {\bar N}}$ and 
$j^{(1v)\, \mu}_{X {\bar N}}$ are shown in Fig.~\ref{figem2}. Note that the 
current $J^{\mu}_{X \Delta}$ is not present because it is set equal to zero 
(see the discussion in Sect.~\ref{kap5_effektiv_d2}). 
In detail, $j^{(1)\, \mu}_{X {\bar N}}$ denotes the 
contact current which is related to the Kroll-Rudermann term of pion 
photoproduction, and $j^{(1v)\, \mu}_{X {\bar N}}$ the vertex-current, 
which is generated by the hadronic vertex form factors. We include in 
$J^{\mu}_{X {\bar N}}(\vec{k})$ in general only the pion 
as most important contribution to the retarded exchange currents with 
one exception, namely those parts which are necessary for generating
the {\it retarded} $\gamma \pi \rho /\omega$ MEC. This current 
is of some relevance in deuteron photodisintegration and therefore is taken 
into account. The remaining current $j^{(0)\, \mu}_{X {\bar N}}$, which is 
a pure mesonic transition current, is shown in diagram (a) of 
Fig.~\ref{figem3} together with the two one-body parts of the diagonal 
component $J^{\mu}_{XX}(\vec{k})$, describing the coupling of a photon to a 
nucleon (meson) with a spectating meson-nucleon (nucleon-nucleon) state 
(see diagrams (b) and (c), respectively).  These contributions are 
denoted as  $j^{{\bar N}\, \mu}_{XX}(\vec{k})$ and 
$j^{X \, \mu}_{XX}(\vec{k})$, respectively.
The current $j^{(0)\, \mu}_{X {\bar N}}$ 
gives an important contribution to the retarded 
MEC which, however, violates the one-meson-approximation, being another 
pathology of this approximation which we had alluded to in I. But
neglecting it, would lead to a severe violation of gauge invariance for the 
retarded $\pi$ MEC. Therefore, it has to be included perturbatively in the 
construction of the corresponding effective exchange operator.

\subsection{Explicit form of the current components}\label{kap5_jnn}

\begin{figure}[ppp]
\centerline{\psfig{figure=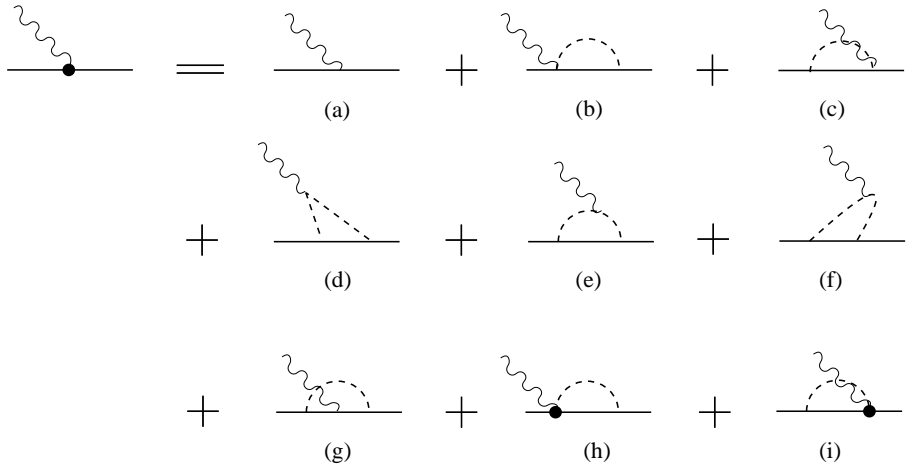,width=12cm,angle=0}}
\vspace{0.5cm}
\caption{
Diagrammatic representation of the separate contributions to 
the {\it physical} one-nucleon current, represented in contrast to the 
bare nucleon current by a filled circle: (a) bare nucleon current, (b) and (c) 
contact current, (d) - (f) meson-in-flight current, (g) bare nucleon current 
with meson in flight, (h) and (i) vertex current.} 
\label{figem4}
\end{figure}

Most of the currents are obtained by the canonical method of minimal coupling, 
and the corresponding matrix elements of ${J}^{\,\mu}$ are derived
in a straightforward manner and need not be given here explicitly. 
The  nucleon one-body current, appearing in 
$j^{[1]\, \mu}_{{\bar N}{\bar N}}$ and 
 $j^{{\bar N}\, \mu}_{XX}(\vec{k})$, needs some special consideration. 
Since we start from bare nucleons, the corresponding current 
$j^{[1]\, \mu}_{{\bar N}{\bar N}}$ is given by a Dirac current with 
vanishing anomalous magnetic moment $\kappa$. However, as discussed in detail 
in I, the bare nucleon becomes dressed by meson-nucleon loop contributions 
(see Appendix A of I). Consequently, the dressed nucleon acquires a 
nonvanishing anomalous magnetic moment by the presence of this meson cloud
 (see Fig.~\ref{figem4}). 
However, it is not in agreement with  the experimental  value as needed 
for a realistic treatment of photonuclear reactions. Therefore,
we substitute the dressed nucleon current of the model by the 
onshell current with the experimental value for $\kappa$.
The corresponding nonrelativistic current is denoted by 
${j}_{real}^{nr,\,\mu}(\vec{k})$ and contains the usual convection and 
spin part. In addition, we also take into account the relativistic 
spin-orbit current as most important relativistic contribution of leading 
order. This relativistic contribution is quite important 
in photodisintegration of the deuteron, even at low 
energies~\cite{CaM82,WiL88}.

It is obvious that the above substitution of the physical onshell nucleon 
current creates formally several inconsistencies. For example, the neglect 
of the explicit evaluation 
of the loop diagrams leads to a violation of unitarity if the nucleon current 
operator is evaluated in the two-nucleon system where the initial and/or 
final nucleon can become offshell. In this case, some of the loops implicitly 
present could become singular above pion threshold and, therefore, would 
lead to an imaginary part describing real meson production. 
%However, this 
%violation can be overcome by explicitly including pion production diagrams.

Now we will consider the remaining current contributions. 
The matrix elements of the vertex current $j^{(1v)\, \lambda}_{{\bar N} X}$
are given by \cite{Sch86}
\begin{eqnarray}
\bra{{\bar N}(\vec{p}^{\,\prime})} \rho^{(1v)}_{{\bar N} X_{\pi}}(\vec{k}) 
 \ket{ \pi(\vec{q},\mu){\bar N}(\vec{p}\,)} &=& 0 \,, \label{kap5_vertex_3}\\
\bra{{\bar N}(\vec{p}^{\,\prime})} 
\vec{\jmath}^{\,\, (1v)}_{{\bar N}X_{\pi}}(\vec{k})
\ket{ \pi(\vec{q},\mu) {\bar N}(\vec{p}\,)} &=&
 \delta(\vec{q} + \vec{p} +\vec{k}- \vec{p}^{\, \prime})\,
 \frac{g^0_{\pi}}{2M_N}\, [\hat{e},  \tau_{\mu} ]\, (2 \vec{q} +  \vec{k})\, 
 i \vec{\sigma} \cdot ( \vec{q}  + \vec{k} )\, \frac{F_{\pi}(\vec{q}^{\,2}) -
 F_{\pi}( (\vec{q}+\vec{k})^{\,2}) }{
      \vec{q}^{\, 2} - (\vec{q}+\vec{k})^{\,2} }  \,,
  \label{kap5_vertex_4}
\end{eqnarray}
where $\hat e= e\,(1+\tau_3)/2$, and 
a pion state with momentum $\vec q$ and isospin projection $\mu$ is 
denoted by $\ket{ \pi(\vec{q},\mu)}$.
As next, we identify the nucleon current contribution in the presence of a 
spectator meson, $j^{{\bar N}\,\mu}_{XX}(\vec{k})$, with the normal nucleon 
current
\begin{eqnarray}
  \bra{ \pi(\vec{q},\mu) {\bar N}(\vec{p}^{\,\prime})}
  j^{{\bar N}\,\lambda}_{XX}(\vec{k})
 \ket{ \pi(\vec{q}^{\,\prime},\mu^{\prime}) {\bar N}(\vec{p}\,)} &=&
 \delta_{\mu \mu^{\prime}}\, \left( 2\pi \right)^3 \, 2 \omega(\vec{q}\,) \,
 \delta(\vec{q}^{\,\prime} - \vec{q}\,) \,
  \bra{{\bar N}(\vec{p}^{\,\prime})}
  {j}_{real}^{nr,\,\lambda}(\vec{k})
 \ket{{\bar N}(\vec{p}\,)}\,. \label{kap5_j_qq_n_modifiziert_2}
\end{eqnarray} 

The current components for the $\gamma\pi\rho/\omega$ currents cannot be 
obtained by minimal substitution, but from the usual interaction 
Lagrangian~\cite{DaM91}
 \begin{equation}\label{kap5_lagrange_vektor}
 {\cal L}_{\gamma \pi x }= \frac{e g_{\gamma \pi x}}{4 m_x}
  \epsilon_{\mu \nu \lambda \sigma} \left(\partial^{\nu} A^{\mu}
  - \partial^{\mu} A^{\nu} \right) \left( \partial^{\sigma} \phi^{\lambda}_x
  - \partial^{\lambda} \phi^{\sigma}_x \right) \phi_{\pi} \quad
  \mbox{with} \quad x \in \{\rho, \omega \} \,\, .
 \end{equation}
The electromagnetic coupling constants $g_{\gamma \pi x}$, as determined from
the decay of the vector meson into a pion and a photon, 
are~\cite{DuK83,Lev95} 
\begin{equation}\label{kap5_zerfall_rho_omega1}
  g_{\gamma \pi \rho} = 0.41 \,\, , \quad  g_{\gamma \pi \omega} = 2.02 \,\, .
\end{equation}

Now, we will turn to the currents in which a $\Delta$ isobar is involved.
For the direct electromagnetic exitation 
$\gamma + {\bar N} \rightarrow \Delta$ we take the usual nonrelativistic form
which contains a magnetic dipole ($M1$) and an electric quadrupole ($E2$)
transition. The analysis of pion photoproduction on the nucleon indicates 
that the $E2$ transition is largely suppressed compared to the dominant 
$M1$ transition. Therefore, we neglect the $E2$ contribution 
and thus take as current~\cite{WeA78}
\begin{eqnarray}
  \bra{\Delta(\vec{p}_{\Delta})} 
 \vec{\jmath}_{\Delta{\bar N}}(\vec k\,) \ket{{\bar N}(\vec{p}_{\bar N})} 
  &=&  e\,\tau_{\Delta {\bar N},\,0}\, 
 \delta(\vec{p}_{\Delta}-\vec{p}_{\bar N}-\vec{k})\,
 \frac{G_{M1}^{0\,\Delta {\bar N}}}{2M_N}
 \,i\vec{\sigma}_{\Delta {\bar N}}\times\vec{k}_{\gamma N}  
\label{ndeltacurrent}\,,
\end{eqnarray}
with
\begin{eqnarray}
\vec{k}_{\gamma N}&=&\vec{k} - 
\frac{M^{res}_{\Delta} - M_N}{M^{res}_{\Delta}}\,\vec{p}_{\Delta}\,,
 \quad \mbox{where} \quad M^{res}_{\Delta} = 1232 \, \mbox{MeV}\,.
\end{eqnarray}
%The term proportional to $\vec{p}_{\Delta}$
%describes a current, associated with a nonvanishing 
%small $\gamma{\bar N} \Delta$ charge contribution, which is neglected in this 
%work. 
%
The corresponding $\gamma{\bar N} \Delta$ charge, which is
  associated with the current contribution  proportional to $\vec{p}_{\Delta}$,
 is of minor importance and therefore neglected in this  work. 
The $M1$ strength is determined by fitting the $M_{1+}^{(3/2)}$ multipole 
amplitude in pion photoproduction on the nucleon which will be the topic of 
the next section.

At the end of this section we will fix the remaining current contributions 
$J^{\mu}_{\Delta \Delta}(\vec{k})$, 
$J^{[2]\,\mu}_{{\bar N}{\bar N}}(\vec{k})$ and 
$J^{[2]\,\mu}_{\Delta{\bar N}}(\vec{k})$. Due to the absence of 
$\Delta \bar N$ components in the deuteron wave function, the current 
$J^{\mu}_{\Delta \Delta}(\vec{k})$ cannot contribute to the 
electromagnetic deuteron break-up.
$J^{[2]\,\mu}_{{\bar N}{\bar N}}(\vec{k})$ is identified with 
the $\rho$ MEC $J^{(\rho)\, \mu}(\vec{k})$, which is the most important 
two-body current besides the  $\pi$ MEC. Note, that first of all in
$J^{[2]\,\mu}_{{\bar N}{\bar N}}(\vec{k})$ as well as in 
$J^{[2]\,\mu}_{\Delta{\bar N}}(\vec{k})$ bare couplings occur, which 
have to be renormalized (see Eq.~\ref{kap5_j_n2_ren_2} in 
Sect.~\ref{kap5_effektiv_n2}). In our approach, $J^{(\rho)\, \mu}(\vec{k})$ 
is taken in the strictly  nonrelativistic form~\cite{ScW90}.
One should note that, due to the neglect of retardation in 
$\vec{J}^{\,\, (\rho)}(\vec{k})$, this MEC is not fully consistent 
with the $\rho$ exchange of the Elster-Potential which serves as basic 
$NN$ interaction in this work (see I). Similarly, the second contribution 
$J^{[2]\,\mu}_{\Delta{\bar N}}(\vec{k})$ is identified with the static 
pionic $\Delta$ MEC $\vec{J}^{\,\, (\pi)}_{\Delta}(\vec{k})$ in the nonrelativistic 
limit~\cite{Wil92}. In photodisintegration, this current contribution is 
almost negligible. 

\section{The $M_{1+}^{(3/2)}$ multipole of pion photoproduction}\label{kap5_m1plus}

\begin{figure}
\centerline{\psfig{figure=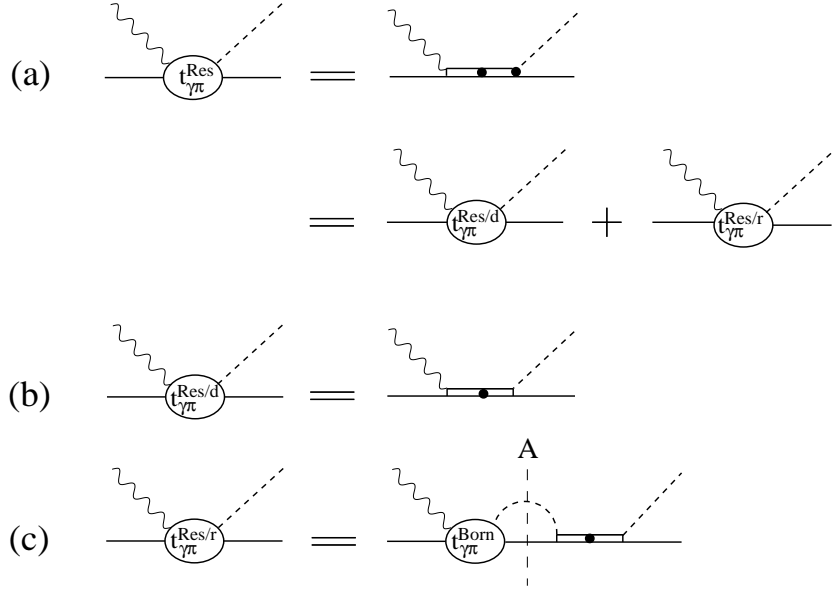,width=11cm,angle=0}}
\vspace{0.5cm}
\caption{
Diagrammatic representation of the resonant amplitude of 
pion photoproduction (a) consisting of a direct amplitude (b),
and a rescattering contribution (c).}
\label{figem7}
\end{figure}

In order to fix the coupling constant $G_{M1}^{0\,\,\Delta {\bar N}}$ 
for the $M1$ excitation of the $\Delta$ resonance, we now will consider 
the $M_{1+}^{(3/2)}$ multipole of pion photoproduction on the nucleon.
The production amplitude is given by the sum of a Born 
and a resonance part (see Fig.~\ref{figem7}) 
\begin{eqnarray}
 t_{\pi \gamma}(z) &=&
   t^{Born}_{\pi \gamma}(z)+
 %  t^{Res/d}_{\pi \gamma}(z) +   t^{Res/r}_{\pi \gamma}(z)
  t^{Res}_{\pi \gamma}(z)  \label{kap5_gamma_pi_2_1}\,.
\end{eqnarray}
The resonance part is given in compact form by
\begin{equation}\label{kap5_kompakt}
 t^{Res}_{\pi \gamma}(z)= 
 v_{X_{\pi} \Delta } 
 g_{\Delta}(z)
 {\widetilde \jmath}_{\Delta {\bar N}}(z) \,,
\end{equation} 
where $g_{\Delta}$ denotes the dressed $\Delta$ propagator as introduced in I. 
The effective, energy dependent $\gamma {\bar N} \Delta$ current
${\widetilde \jmath}_{\Delta {\bar N}}(z)$ contains a direct and a 
rescattering part, i.e.\
\begin{equation}\label{kap5_effektiv}
 {\widetilde \jmath}_{\Delta {\bar N}}(z) =
 \frac{{j}_{\Delta {\bar N} }}{{\bar N}_{[1]}} +
  v_{\Delta X_{\pi}} g_0(z)
  t^{Born}_{\pi \gamma}(z) \,.
 \end{equation}
where $g_{0}$ denotes the nucleon propagator, and 
${\bar N}_{[1]}$ a renormalization constant in the one-nucleon
sector as defined in the Appendix A of I.  

\begin{figure}[t]
\centerline{\psfig{figure=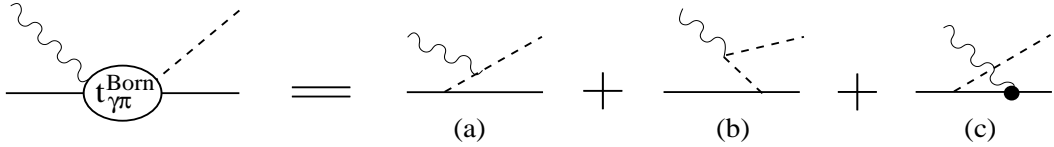,width=14cm,angle=0}}
\vspace{0.5cm}
\caption{Diagrammatic representation of the Born amplitude
$t^{Born}_{\pi \gamma\, \mu}(z)$
 in the $M_{1+}(3/2)$-multipole amplitude.
The separate diagrams are denoted as follows:
time ordered pion pole terms (a) and (b) and crossed nucleon pole
 term (c).}
\label{figem6}
\end{figure}

\begin{figure}[btp]
\centerline{\psfig{figure=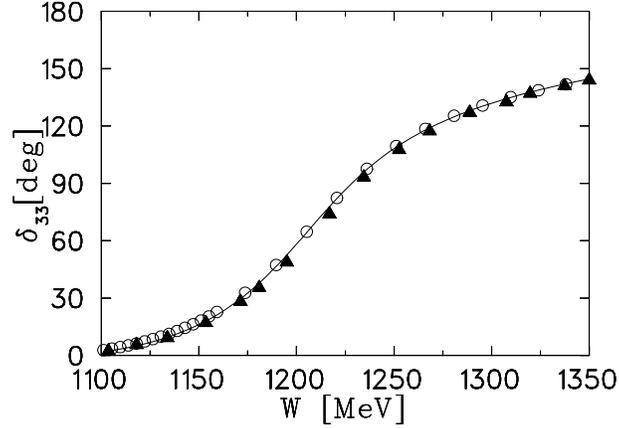,width=8cm,angle=90}}
\vspace{0.5cm}
\caption{Experimental $P_{33}$-scattering phase $\delta_{33}$ of $\pi N$ 
scattering (triangles from~\protect\cite{Arn98}, solution SM95) and  
photoproduction phase $\delta_{M_{1+}^{(3/2)}}$ 
(open circles from~\protect\cite{Arn98}, solution SM97K). The solid curve 
represents our fit using the values of Eq.~(67) in I, in which
both data sets are considered with equal weight.}
 \label{figem8}
\end{figure}

\begin{figure}[btp]
\centerline{\psfig{figure=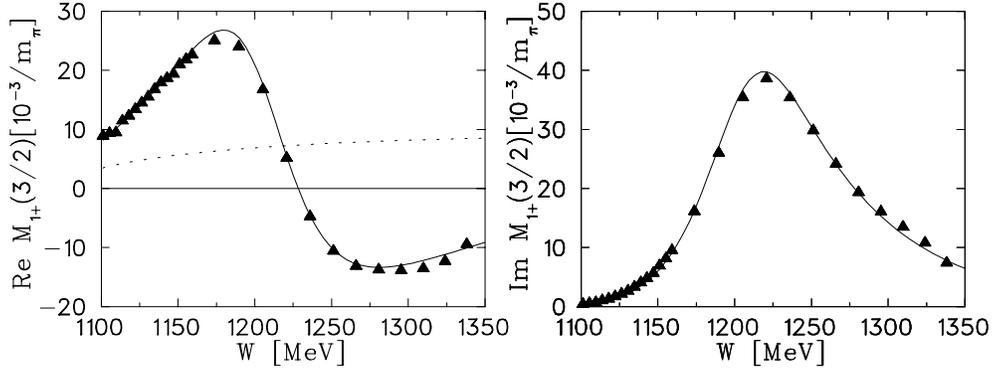,width=13cm,angle=90}}
\vspace{0.5cm}
\caption{Fit of the real and the imaginary part of the
$M_{1+}^{(3/2)}$ multipole amplitude to the experimental data from
\protect\cite{Arn98}, solution SM97K. The dotted curve represents the
contribution of the Born amplitude.}
\label{figem9}
\end{figure}

The dominant contributions to the Born amplitude to the $M_{1+}^{(3/2)}$ 
amplitude are displayed in Fig.~\ref{figem6}. They consist of the time 
ordered pion pole diagrams and the  crossed  nucleon 
pole diagram. Following 
the procedure in~\cite{TaO85}, less important mechanisms like 
$\omega$ exchange will be taken into account effectively by a parameter 
$b$ which is close to unity. Thus we use explicitly as Born amplitude
\begin{eqnarray}
 \bra{\pi(\vec{q}\,\mu) N(-\vec{q}\,)}
 t^{Born}_{M_{1+}^{(3/2)}}(z)\ket{ \gamma(\vec{k}\,)   
 {\bar N}(-\vec{k}\,)} &=& 
- b\,\frac{g_{\pi}}{2M} [\hat{e},\tau_{\mu}^{\dagger}]\,
  \vec{\epsilon}\!\cdot\!(2\vec{q}-\vec{k})\,
 i\vec{\sigma}\!\cdot\!(\vec{q}-\vec{k}) \, F_{\pi}((\vec{q} - \vec{k})^{\,2}) 
  \nonumber \\
  && \times \frac{1}{2\omega(\vec{q}-\vec{k})}\, 
  \left[
   \frac{1}{z-k-\omega(\vec{q}-\vec{k}\,)-e^{nr}_N(\vec{q}\,)}
  + \frac{1}{z-\omega(\vec{q}\,)-\omega(\vec{q}-\vec{k})-e^{nr}_N(\vec{k})}
  \right]\,\,   \nonumber \\ 
  & &
 -b\,\frac{g_{\pi}}{4M_N^2}
  \frac{\hat{e}\tau_{\mu}^{\dagger}\,
  \vec{\epsilon}\!\cdot\!(2\vec{q}+\vec{k})\,i\vpr{\sigma}{q}
  +(\hat{e}+\hat{\kappa})\tau_{\mu}^{\dagger}\,
  \vec{\epsilon}\!\cdot\!\vec{\sigma}\times\vec{k}\,\vpr{\sigma}{q}}
  {z-k-\omega(\vec{q}\,)-e^{nr}_N(\vec{q}+\vec{k})}\,
  F_{\pi}(\vec{q}^{\,2}) \,,
 \label{kap5_born_explizit_1}
\end{eqnarray}
where the form factor $F_{\pi}$ and the pion-nucleon coupling constant
$g_{\pi}$ are given by the underlying $NN$ interaction (here the 
Elster-potential, see Table I in I). The parameter $b$ and the renormalized 
$\gamma {\bar N} \Delta$ coupling constant 
$G_{M1}^{\Delta {\bar N}}=G_{M1}^{0\,\,\Delta {\bar N}}/{\bar N}_{[1]}$
are fitted to the experimental data of Arndt {\it et al.}~\cite{Arn98} 
(solution SM97K) yielding
\begin{equation}\label{kap5_chi}
 G_{M1}^{\Delta {\bar N}} = -3.91\,\,\mbox{ and }\,\, b= 0.84\,.
\end{equation}
We would like to emphasize that Watson's theorem~\cite{Wat54} is exactly
fulfilled below two-pion threshold. In view of small but significant
differences of the experimental $P_{33}$-phases determined from
$\pi N$ scattering versus pion photoproduction, 
we have determined $G_{M1}^{\Delta {\bar N}}$ and $b$ as well as
the hadronic parameters 
$f^0_{\Delta \pi N}$, $\Lambda_{\Delta \pi N}$, and $M^0_{\Delta}$ by a 
{\it simultaneous} fit of the $M_{1+}^{(3/2)}$ multipole and the $P_{33}$ 
partial wave of $\pi N$ scattering (see Figs.~\ref{figem8} 
and \ref{figem9}).

\begin{figure}[ttt]
\centerline{\psfig{figure=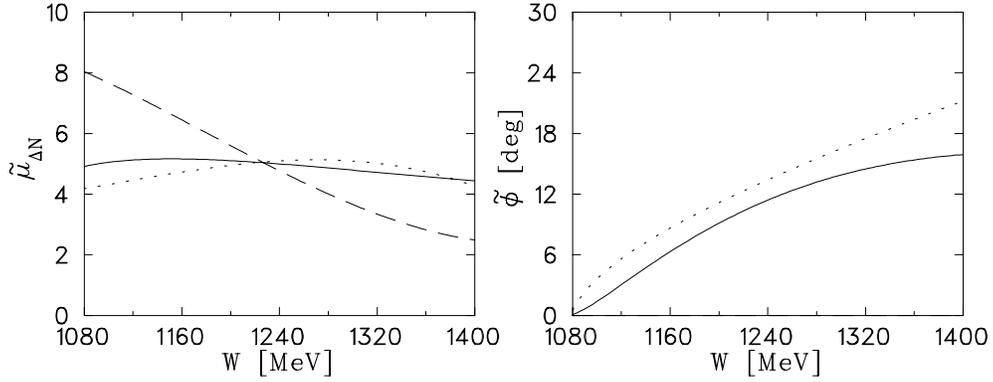,width=13cm,angle=90}}
\vspace{0.5cm}
\caption{Energy dependence of the effective $\gamma {\bar N} \Delta$ coupling
for the onshell-case $W= e^{nr}_N(\vec{k}) + k$. The left panel shows 
the modulus $\widetilde{\mu}_{\Delta {\bar N}}(W,k)$ and the right panel 
the phase $\widetilde{\Phi}(W,k)$. The full curve represents the coupling 
$\widetilde{G}_{M1}^{\Delta {\bar N}}(\mbox{eff1})$ in 
(\protect\ref{kap5_effektiv}), and the dotted curve the one of Wilhelm 
$\widetilde{G}_{M1}^{\Delta {\bar N}}(\mbox{eff2})$
(see (\protect\ref{kap5_parametrisierung_3})). The dashed curve shows the
``modified'' coupling of Wilhelm {\it et al.} 
$\widetilde{G}_{M1}^{\Delta {\bar N}}(\mbox{eff3})$
 (see the discussion in
 Sect.~\ref{kap5_stat_inkon}).
} 
 \label{figem10}
\end{figure}

As next, we would like to introduce the effective strength
${\widetilde G}_{M1}^{\Delta {\bar N}}(z,k)$ which is defined by the 
matrix element of the effective current 
${\widetilde \jmath}^{\,\mu}_{\Delta {\bar N}}(z)$ in the $\Delta$ rest frame  
\begin{equation}\label{kap5_j_effektiv_2}
   \bra{\Delta(\vec{p}_{\Delta}=0)}
 \vec{ {\widetilde \jmath}}_{\Delta {\bar N}}(z,\vec{k})
 \ket{{\bar N}(\vec{p}_{\bar N})} =
   e\,\tau_{\Delta {\bar N},\,0}\,
 \delta(\vec{p}_{\bar N} + \vec{k} ) 
  \, \frac{{\widetilde G}_{M1}^{\Delta {\bar N}}(z,k)}{2M_N} 
 \,i\vec{\sigma}_{\Delta {\bar N}}\times\vec{k} \,\, ,
\end{equation} 
where for initial onshell nucleons $z$ and k are related by 
$z= W + i \epsilon$ with $W= e^{nr}_N(\vec{k}) + k$. In the later 
discussion of the results this coupling will be referred to as 
$\widetilde{G}_{M1}^{\Delta {\bar N}}(\mbox{eff1})$. Because of the 
occurrence of intermediate pion-nucleon loops in the rescattering amplitude 
(see Fig.~\ref{figem7}), ${\widetilde G}_{M1}^{\Delta {\bar N}}(z,k)$ 
becomes complex above $\pi$ threshold. Modulus 
$\widetilde{\mu}_{\Delta N}(W,k)$ and 
phase $\widetilde{\Phi}(W,k)$ of the effective coupling
\begin{equation}\label{kap5_parametrisierung_1}
   \widetilde{G}_{M1}^{\Delta {\bar N}}(z=W+i \epsilon,k)=
  \widetilde{\mu}_{\Delta N}(W,k)\,e^{i\widetilde{\Phi}(W,k)}
\end{equation}
are shown by the full curves in Fig.~\ref{figem10}.
Obviously, the energy dependence of the modulus is rather weak. 

Finally, we would like to compare briefly the effective 
$\gamma {\bar N} \Delta$ coupling of Wilhelm {\it et al.}~\cite{Wil92,WiA93}. 
Similar to \cite{KoM83,KoM84}, the onshell matrix element of the resonant 
amplitude together with the Born terms are 
fitted directly to experimental data without explicitly evaluating the 
rescattering amplitude. The parameters $b$ and $F_{\pi}$ are set equal 
to one, and the energy dependent ansatz is used
\begin{eqnarray}
 \widetilde{\mu}_{\Delta N}(W,k)=
 \mu_0+\mu_2\left(\frac{q}{m_{\pi}}\right)^2
    +\mu_4\left(\frac{q}{m_{\pi}}\right)^4\,\mbox{ and }
     \,\, \widetilde{\Phi}(W,k)=\frac{q^3}{a_1+a_2q^2}\,,
   \label{kap5_parametrisierung_3} 
\end{eqnarray}
where $W= e^{nr}_N(\vec{q}\,) + \omega_{\pi}(\vec{q}\,)$ and  
$q$ denotes the momentum of the outgoing pion. The free parameters are fitted 
to the data of Berends and Donnachie~\cite{BeD75}. 
This coupling, denoted henceforth by 
$\widetilde{G}_{M1}^{\Delta {\bar N}}(\mbox{eff2})$, 
is  represented by the dotted curves in Fig.~\ref{figem10}. Note that 
the modulus of this coupling is considerably smaller at low energies than 
the one of our approach, where the rescattering 
amplitude is explicitly evaluated. As will be seen in 
Sect.~\ref{kap6_deutspalt}, this feature turns out to be one of the reasons 
for the considerable underestimation of the total cross section of deuteron 
photodisintegration in \cite{Wil92,WiA93}.

\section{The effective current operator in deuteron photodisintegration}
\label{kap5_effektiv_strom}

After having fixed all current components of (\ref{kap5_strom_matrix}), 
we will now construct an effective current operator, which acts in pure 
hadronic space, by eliminating all explicit meson d.o.f. It is defined by
\begin{equation}\label{kap5_effektiv_def}
 \bra{f}P J_{eff}^{\mu}(z,\vec{k}) P \ket{i}
  = \bra{f} J^{\mu}(\vec{k}) \ket{i} \,.
 \end{equation}  
The initial and final states are given by 
the effective deuteron and outgoing $NN$ scattering states, respectively.
Their explicit forms are given in (55) and (47) through (49) of I.
We will evaluate the effective current in the antilab system, where the final 
$NN$ scattering state is at rest and the deuteron has total momentum $-\vec k$.
In principle, one would then need a boost contribution for the moving 
deuteron which, however, can safely be neglected according to the findings 
in~\cite{GoA92}.

\begin{figure}[tp]
\centerline{\psfig{figure=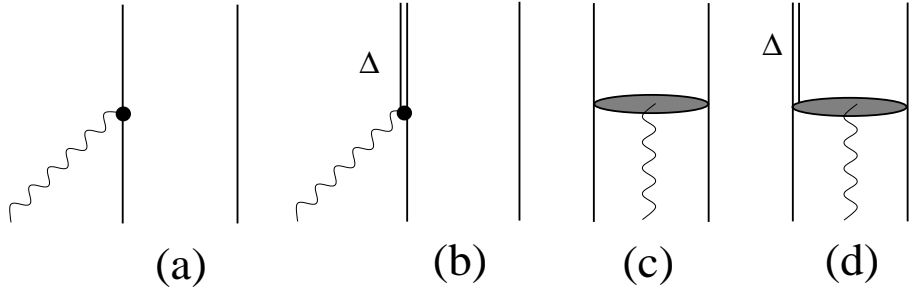,width=12cm,angle=0}}
\vspace{0.5cm}
\caption{Diagrammatic representation of the separate contributions
to the effective current operator $J_{eff}^{\mu}(z,\vec{k})$ (see
Eq.~(\ref{kap5_eff_zerlegung})):
 (a) $J_{eff}^{N[1]\, \mu}(z,\vec{k})$,
 (b) $J_{eff}^{\Delta[1]\, \mu}(z,\vec{k})$,
 (c) $J_{eff}^{N[2]\, \mu}(z,\vec{k})$, and
 (d) $J_{eff}^{\Delta[2]\, \mu}(z,\vec{k})$.}
\label{figem11}
\end{figure}

Inserting the various components of the current into the rhs of 
(\ref{kap5_effektiv_def}), it turns out that the effective current operator
can be split into a nucleonic and a resonant part with superscripts ``$N$'' 
and ``$\Delta$'', respectively, which in turn can be divided into one- and 
two-body terms denoted by superscripts ``$[1]$'' and ``$[2]$'', respectively 
(see Fig.~\ref{figem11})
\begin{equation}\label{kap5_eff_zerlegung}
 J_{eff}^{\mu}(z,\vec{k}) =
 J_{eff}^{N[1]\, \mu}(z,\vec{k}) +
 J_{eff}^{\Delta[1]\, \mu}(z,\vec{k}) +
 J_{eff}^{N[2]\, \mu}(z,\vec{k}) +
 J_{eff}^{\Delta[2]\, \mu}(z,\vec{k}) \, .
\end{equation}
We would like to remark, that in this evaluation we have included 
perturbatively the contributions from the one-meson-approximation violating 
diagram (a) of Fig.~\ref{figem3}. In this context, we have to introduce
into the corresponding expressions below projection operators 
$P_{x_1 x_2}$ on a state consisting of two nucleons and two mesons of type 
$x_1$ and $x_2$. 

\subsection{The nucleonic one-body contribution 
$J_{eff}^{N[1]\, \mu}(z,\vec{k})$}\label{kap5_effektiv_n1}

According to the discussion in subsection \ref{kap5_jnn}, this component 
is represented by the physical onshell current including the relativistic 
spin-orbit part
\begin{eqnarray}
 {^{(-)}\bra{NN;\,\vec{p}, \alpha)}}P 
  J^{N[1]\, \mu}_{eff}(z,\vec{k}) P \ket{d}
 &=& \frac{1}{N_d}  
  {^{(-)}\bra{NN;\,\vec{p}, \alpha)}}P_{\bar N}
 \frac{{\widehat Z}^{os}_{[2]}}{{\widehat R}(z)}
 \sum_{i=1,2}  \Big( 
 j_{real}^{nr,\, \mu}(i,\vec{k}) + j^{so\, \mu}(i,\vec{k})  \Big)
  P_{\bar N} \ket{d}\,,
  \label{kap5_eff_1_ende}
 \end{eqnarray}
where $\alpha$ characterizes additional quantum numbers, and 
the normalization constant $N_d$ is given in Eq.~(56) of I. It differs
from unity in retarded approaches only (e.g., $N_d = 0.992$ for the 
Elster-potential).

\subsection{The resonant one-body contribution 
$J_{eff}^{\Delta[1]\, \mu}(z,\vec{k})$}\label{kap5_effektiv_d1}

The component $J_{eff}^{\Delta[1]\, \mu}(z,\vec{k})$ is given essentially 
by the effective $\gamma {\bar N} \Delta$ current 
\begin{eqnarray}
 {^{(-)}\bra{NN;\,\vec{p}, \alpha)}}P_{\Delta} 
 J^{\Delta[1]\, \mu}_{eff}(z,\vec{k})
  P_{\bar N}\ket{d} &=& \frac{1}{N_d}
 {^{(-)}\bra{NN;\,\vec{p}, \alpha)}}P_{\Delta}
 \sum_{i,j=1,2;\,i\neq j} 
  \Big({\widetilde {\jmath}}^{\,\mu}_{\Delta{\bar N}}(i,z -
  e_N(\vec{p}_j),\vec{k})\Big)
  {\hat R}^{\pi}(z-k) P_{\bar N} \ket{d} \,.
  \label{kap5_eff_nd_1}
\end{eqnarray}
Note that similar to the Elster-Potential, we consider only pion-nucleon
loops in ${\hat R}^{\pi}(z-k)$. The effective current 
${\widetilde \jmath}^{\,\mu}_{\Delta {\bar N}}(z)$ has the same functional 
form as given in (\ref{ndeltacurrent}) except that 
$G_{M1}^{0\,\Delta {\bar N}}$ is replaced be the effective excitation strength 
${\widetilde G}_{M1}^{\Delta {\bar N}}(W_{sub}+ i \epsilon, k_{\gamma N})$,
where
\begin{eqnarray}
   W_{sub}(W,\vec{p}_{\Delta})
  &=& W-M_N- \frac{\vec{p}_{\Delta}^{\,2}}{2\mu_{N\Delta}}\, 
 \label{kap5_k_wsub}
\end{eqnarray}
denotes the invariant mass of the $\pi N$ subsystem which we have evaluated 
for the spectator onshell choice. In the actual evaluations in the two-nucleon 
system, ${\widetilde G}_{M1}^{\Delta {\bar N}}$ is 
calculated in the rest frame of the $\Delta$, similar to the treatment
in the $M_{1+}^{(3/2)}$ multipole. The difference
 with respect to the one-nucleon sector is that $k_{\gamma N}$ and 
$W_{sub}$ are {\it not} related by an onshell relation, i.e., 
$e^{nr}_N(k_{\gamma N}) + k_{\gamma N} \neq W_{sub}$. Because of the 
explicit evaluation of the pion-nucleon loop in the rescattering
Born amplitude (cut A in diagram (c) of Fig.~\ref{figem7}), this feature does 
not lead to any complications in practice. On the other hand, in the 
treatment of Wilhelm {\it et al.}~\cite{WiA93} such an offshell extrapolation 
of ${\widetilde G}_{M1}^{\Delta {\bar N}}$ is not possible. Therefore, they 
have used an onshell prescription
\begin{equation}
  \widetilde{G}_{M1}^{\Delta {\bar N}}(W_{sub}+ i \epsilon,
 k_{\gamma N}) \to 
  \left\{ \begin{array}{ll}
   {G}_{M1}^{\Delta N}(E_\Delta) &
  \mbox{for} \quad E_\Delta=W_{sub} >m_{\pi}+M_N\, ,\\
   {G}_{M1}^{\Delta N}(M_N + m_{\pi}) & \mbox{otherwise}\, .
  \end{array} \right. 
  \label{kap5_paul_on}
\end{equation}
It turns out that this onshell 
treatment for ${\widetilde G}_{M1}^{\Delta {\bar N}}$ is a very good 
approximation, at least for photodisintegration, compared to the exact 
calculation, where the arguments $k_{\gamma N}$ and $W_{sub}$ in 
(\ref{kap5_paul_on}) are treated independently.

\subsection{The nucleonic meson exchange current 
$J_{eff}^{N[2]\, \mu}(z,\vec{k})$}\label{kap5_effektiv_n2}

\begin{figure}[ppp]
\centerline{\psfig{figure=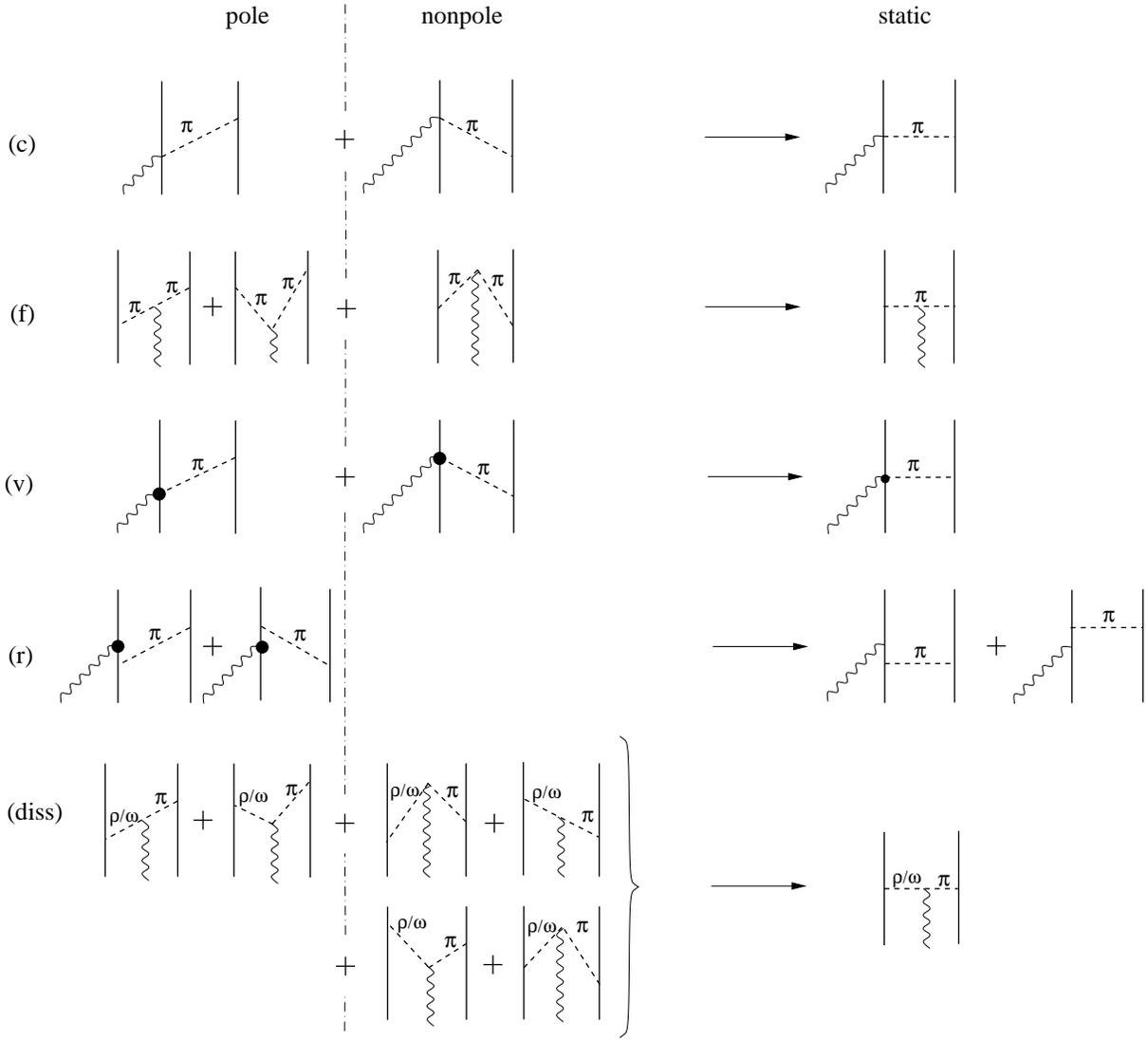,width=16cm,angle=0}}
\vspace{0.5cm}
\caption{Diagrammatic representation of the retarded $\pi$ (contact (c), 
pion-in-flight (f), vertex (v), recoil (r)) and $\gamma \pi\rho /\omega$ 
MEC (diss). The arrows indicate the static limit.}
 \label{figem13}
\end{figure}

\begin{figure}[btp]
\centerline{\psfig{figure=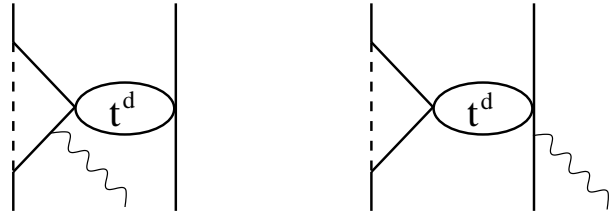,width=8cm,angle=0}}
\vspace{0.5cm}
\caption{Examples for effective exchange currents, which are neglected due to the
substitution (\ref{Gsubstitute}).}
 \label{figem12}
\end{figure}

The matrix element of $J_{eff}^{N[2]\, \mu}(z,\vec{k})$ has the following 
structure 
\begin{eqnarray}
 {^{(-)}\bra{NN;\,\vec{p}, \alpha)}}P_{\bar N} 
 J^{N[2]\, \mu}_{eff}(z,\vec{k}) P_{\bar N} \ket{d}
 &=&  \frac{1}{N_d}
 {^{(-)}\bra{NN;\,\vec{p}, \alpha)}}P_{\bar N}
 J^{N[2]\, \mu}_{eff}(z,\vec{k})
     \frac{{\hat R}(z-k)}{{\hat Z}^{os}_{[2]} } 
 P_{\bar N} \ket{d}\,, \label{kap5_eff_matrix_m2}
\end{eqnarray} 
where $J_{eff}^{N[2]\,\mu}(z,\vec{k})$ consists of a static $\rho$ MEC, 
a retarded $\pi$ MEC, and a retarded $\gamma\pi\rho /\omega$-current 
$J^{(diss)\, \mu}(z,\vec{k}\,)$
\begin{eqnarray}
  J_{eff}^{N[2]\, \mu}(z,\vec{k}) &=& J^{(\rho)\, \mu}(\vec{k}) +
 J^{(\pi)\, \mu}(z,\vec{k}\,)  + J^{(diss)\, \mu}(z,\vec{k}\,)
  \,. \label{kap5_effektiv_n2_expl_1}
\end{eqnarray}
A new feature of retardation is the fact, that above pion threshold
the retarded currents can be split into a part containing a pole 
and a regular part ($nonpole$)
 \begin{eqnarray}
 J^{(\pi)/(diss)\, \mu}(z,\vec{k}\,) &=&
 J^{(\pi)/(diss)\, \mu}_{pole}(z,\vec{k}\,) +
  J^{(\pi)/(diss)\, \mu}_{nonpole}(z,\vec{k}\,)\,.
 \label{kap5_pol_nonpol_2}
 \end{eqnarray}
The pole part describes the reabsorption of a real pion by a nucleon. 
Furthermore, the retarded $\pi$ MEC splits  into a contact ($c$), 
pion-in-flight ($f$), vertex ($v$) and recoil contribution ($r$)
\begin{eqnarray}
 J^{(\pi)\, \mu}(z,\vec{k}\,) &=&
 J^{(\pi)/c\, \mu}(z,\vec{k}\,) + J^{(\pi)/f\, \mu}(z,\vec{k}\,) +
 J^{(\pi)/v\, \mu}(z,\vec{k}\,) + J^{(\pi)/r\, \mu}(z,\vec{k}\,) 
   \, . \nonumber\\  \label{kap5_effektiv_n2_expl_2}
 \end{eqnarray}
A graphical illustration of these currents is given in Fig.~\ref{figem13}. 
In addition, because of the nonvanishing $\pi d$ interaction $V^0_{XX}$,  
currents like those depicted in Fig.~\ref{figem12} would have to be taken into 
account. However, a detailed analysis has shown that these contributions can 
be neglected in deuteron photodisintegration. This approximation 
is realized by the substitution 
\begin{eqnarray}
G^X(z) =\left( z - H_{0,XX} - V^0_{XX} \right)^{-1}
 &\longrightarrow&\,\, 
G_0(z) = \left( z - H_{0,XX}\right)^{-1}
\label{Gsubstitute} 
\end{eqnarray}
in the above expressions (see Eq.~(49) of I). 

Explicitly, one obtains for the various components 
\begin{eqnarray}
 J^{(\pi)/c\, \mu}_{pole}(z,\vec{k}\,) &=& \sum_{i,j=1,2 ;\,j \neq i}
  v^{0}_{{\bar N}X_{\pi}}(j) G_0(z)
  \, j^{(1)\, \mu}_{X_{\pi} {\bar N}}(\vec{k},i)\, ,
 \label{kap5_kontakt_1}\\
 J^{(\pi)/c\, \mu}_{nonpole}(z,\vec{k}\,) &=& \sum_{i,j=1,2 ;\,j \neq i}
   \, j^{(1)\, \mu}_{{\bar N} X_{\pi}} (\vec{k},i)\,
 G_0(z-k) v^{0}_{X_{\pi} {\bar N}}(j)  \, , 
 \label{kap5_kontakt_2}\\
 J^{(\pi)/f\, \mu}_{pole}(z,\vec{k}\,) &=& \sum_{i,j=1,2 ;\,j \neq i}
 \left( 
    v^{0}_{{\bar N}X_{\pi}}(j) G_0(z)
  \, j^{(\pi)\, \mu}_{X_{\pi} X_{\pi}} (\vec{k})\,
  G_0(z-k) v^{0}_{X_{\pi} {\bar N}}(i) \right. \nonumber\\
  & & \quad \quad \quad \quad \left.
  +  v^{0}_{{\bar N} X_{\pi}}(j) G_0(z) v^0_{X_{\pi} X_{\pi \pi}}(i)  G_0(z)
  \, j^{(0)\, \mu}_{X_{\pi \pi} {\bar N} } (\vec{k}) \right)\, ,
 \label{kap5_flug_1}\\
 J^{(\pi)/f\, \mu}_{nonpole}(z,\vec{k}\,) &=& \sum_{i,j=1,2 ;\,j \neq i}
   \, j^{(0)\, \mu}_{{\bar N} X_{\pi \pi}} (\vec{k})\,
 G_0(z-k)
  v^{0}_{X_{\pi \pi}X_{\pi}}(i) G_0(z-k) v^{0}_{X_{\pi} {\bar N}}(j)\, ,
 \label{kap5_flug_2}\\
 J^{(\pi)/v\, \mu}_{pole}(z,\vec{k}\,) &=& \sum_{i,j=1,2 ;\,j \neq i}
  v^{0}_{{\bar N}X_{\pi}}(j) G_0(z)
  \, j^{(1v)\, \mu}_{X_{\pi} {\bar N}}(\vec{k},i)\, ,
 \label{kap5_vertex_c_1}\\
 J^{(\pi)/v\, \mu}_{nonpole}(z,\vec{k}\,) &=& \sum_{i,j=1,2 ;\,j \neq i}
   \, j^{(1v)\, \mu}_{{\bar N} X_{\pi}} (\vec{k},i)\,
 G_0(z-k) v^{0}_{X_{\pi} {\bar N}}(j)  \, , 
 \label{kap5_vertex_c_2}\\
 J^{(\pi)/r\, \mu}_{pole}(z,\vec{k}\,) &=& \sum_{i,j=1,2 ;\,j \neq i}
  \left( 
  v^{0}_{{\bar N}X_{\pi}}(j) G_0(z)
  \, j^{nr,\, \mu}_{real}(\vec{k},i) 
 v^{0}_{X_{\pi} {\bar N}}(i)  + v^{0}_{{\bar N}X_{\pi}}(i) G_0(z)
  \, j^{nr,\, \mu}_{real}(\vec{k},i) 
 v^{0}_{X_{\pi} {\bar N}}(j) \right)
 \, ,
 \label{kap5_rueck_1}\\
 J^{(\pi)/r\, \mu}_{nonpole}(z,\vec{k}\,) &=& 0\, ,
 \label{kap5_rueck_2}\\
 J^{(diss)\, \mu}_{pole}(z,\vec{k}\,) &=& \sum_{i,j=1,2 ;\,j \neq i}
 \Big( v^{0}_{{\bar N}X_{\pi}}(i) G_0(z)
  \, j^{(\pi)\, \mu}_{X_{\pi} X_{x}} (\vec{k})\,
  G_0(z-k) v^{0}_{X_{x} {\bar N}}(j)  \nonumber\\
  & & \quad \quad \quad \quad 
  +  v^{0}_{{\bar N} X_{\pi}}(j) G_0(z) v^0_{X_{\pi} X_{\pi x}}(i)  G_0(z)
  \, j^{(0)\, \mu}_{X_{\pi x} {\bar N} } (\vec{k}) \Big)
\Big|_{x \in\{\rho, \omega \}}\,, 
 \label{kap5_diss_1}\\
 J^{(diss)\, \mu}_{nonpole}(z,\vec{k}\,) &=& \sum_{i,j=1,2 ;\,j \neq i}
 \Big( j^{(0)\, \mu}_{{\bar N} X_{\pi x}} (\vec{k})\,
 G_0(z-k)
  v^{0}_{X_{\pi x}X_{x}}(j) G_0(z-k) v^{0}_{X_{x} {\bar N}}(i) \nonumber\\
 & & \quad \quad \quad \quad 
 + v^{0}_{{\bar N}X_{x}}(i) G_0(z)
  \, j^{(\pi)\, \mu}_{X_{x} X_{\pi}} (\vec{k})\,
  G_0(z-k) v^{0}_{X_{\pi} {\bar N}}(j)  \nonumber\\
  & &  \quad \quad \quad \quad
  +  v^{0}_{{\bar N} X_{x}}(i) G_0(z) v^0_{X_{x} X_{\pi x}}(j)  G_0(z)
  \, j^{(0)\, \mu}_{X_{\pi x} {\bar N} } (\vec{k})  \nonumber\\
  & &  \quad \quad \quad \quad 
  +  j^{(0)\, \mu}_{{\bar N} X_{\pi x}} (\vec{k})\,
 G_0(z-k)
  v^{0}_{X_{\pi x}X_{\pi}}(j) G_0(z-k) v^{0}_{X_{\pi} {\bar N}}(i)
   \Big)\Big|_{ x \in\{\rho, \omega \}}\,.  
 \label{kap5_diss_2}
\end{eqnarray}

In the present evaluation, we have used a 
nonrelativistic reduction of the $\pi {\bar N}$ vertex for the sake of 
simplicity. The bare couplings have to be renormalized, of course, with the 
renormalization operator ${\hat Z}^{os}_{[2]}$ in
(\ref{kap5_eff_matrix_m2}). Therefore, it is useful to introduce
renormalized MECs according to
\begin{eqnarray}
 {\cal J}^{(\rho)/(\pi)/(diss)\, \mu}(z,\vec{k}\,)   &=&
 \frac{1}{{\hat Z}^{os}_{[2]} }
 J^{(\rho)/(\pi)/(diss)\, \mu}(z,\vec{k}\,)   
 \frac{1}{{\hat Z}^{os}_{[2]} }\,,
 \label{kap5_j_n2_ren_2} 
\end{eqnarray}
so that the relevant matrix element (\ref{kap5_eff_matrix_m2}) has the 
structure 
\begin{eqnarray}
 {^{(-)}\bra{NN;\,\vec{p},\alpha}}P J^{N[2]\, \mu}_{eff}(z,\vec{k}) P \ket{d}
 &=& \frac{1}{N_d}
 {^{(-)}\bra{NN;\,\vec{p},\alpha}}P_{\bar N}
\frac{{\widehat Z}^{os}_{[2]}}{{\widehat R}(z)}
 \Big(    {\cal J}^{(\rho)\, \mu}(\vec{k})
 \nonumber\\
 & &  + {\hat R}^{\pi}(z)  \left({\cal J}^{(\pi)\, \mu}(z,\vec{k}\,)  
  +{\cal J}^{(diss)\, \mu}(z,\vec{k}\,)  \right)  {\hat R}^{\pi}(z-k)
  \Big)  P_{\bar N} \ket{d'}
  \,,  \label{kap5_eff_matrix_m3}
 \end{eqnarray} 
where the dressing operator ${\hat R}^{\pi}$ is neglected in the 
$\rho$ MEC due to its minor importance. 

Summarizing the various contributions to the nucleonic meson exchange 
currents, we include as retarded currents the $\pi$ and 
$\gamma \pi \rho /\omega$ MEC, whereas of the heavier mesons only the 
$\rho$ MEC will be retained in the static limit. 
All the other remaining $\sigma$, $\delta$, $\omega$, and $\eta$ MECs 
are completely neglected in view of their negligible contributions found 
previously in~\cite{RiA98}. Moreover, due to the use of Siegert 
operators~\cite{ArS91} and the much larger mass of the heavy mesons, the 
role of the corresponding MEC is largely suppressed.

\subsection{The resonant meson exchange current 
$J_{eff}^{\Delta[2]\, \mu}(z,\vec{k})$}
\label{kap5_effektiv_d2}

\begin{figure}[tbp]
\centerline{\psfig{figure=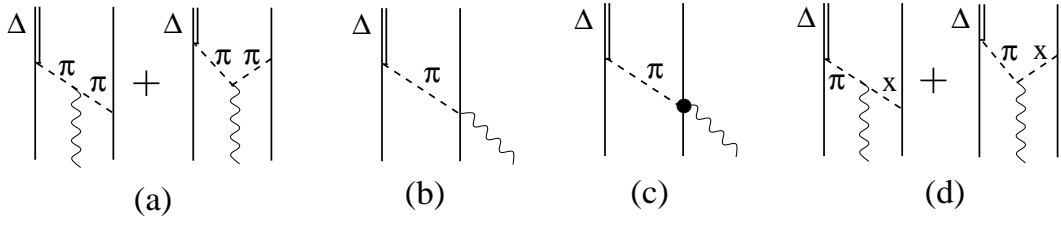,width=14cm,angle=0}}
\vspace{0.5cm}
\caption{Diagrammatic representation of the retarded $\Delta$ MEC: (a)
 meson-in-flight, (b) contact, (c) vertex, and (d) dissociation contributions
($x\in \{\rho,\omega\}$).}
\label{figem14}
\end{figure}

\begin{figure}[t]
\centerline{\psfig{figure=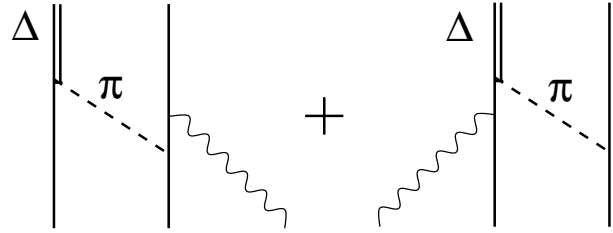,width=8cm,angle=0}}
\vspace{0.5cm}
\caption{Diagrammatic representation of the retarded $\Delta$ recoil MEC
 $J^{(\pi)/r\, \mu}_{\Delta}(z,\vec{k}\,)$.}
\label{figem15}
\end{figure}

In analogy to the nucleonic MEC, we construct the retarded $\Delta$ MEC 
which is depicted in Figs.~\ref{figem14} and \ref{figem15}. The currents 
represented in Fig.~\ref{figem14} are the contact, meson-in-flight, vertex, 
and dissociation contributions to the $\Delta$ MEC. In the static limit, 
they are already taken into account via the static $\Delta$ MEC (apart from 
the dissociation current) so that the problem of double counting arises. In 
view of the fact that this static $\Delta$ MEC leads to a very small 
contribution in deuteron photodisintegration, it is resonable to assume that 
retardation will not change this fact qualitatively. Therefore, we take 
the diagrams of Fig.~\ref{figem14} in the static limit as already contained in 
the static $\Delta$ MEC and consequently set 
$J^{[2]\,\mu}_{X \Delta}(\vec{k})$ equal to zero. 

The recoil contribution of Fig.~\ref{figem15}, on the other hand, 
is not contained in $J^{[2]\,\mu}_{\Delta{\bar N}}(\vec{k})$. Therefore, 
it is explicitly taken into account. Introducing again 
renormalized MECs, we obtain  for the matrix element of the $\Delta$ MEC
\begin{eqnarray}
 {^{(-)}\bra{NN;\,\vec{p},\alpha}}P_{\Delta} 
 J^{\Delta[2]\, \mu}_{eff}(z,\vec{k}) P_{\bar N} \ket{d}
 &=& \frac{1}{N_d} {^{(-)}\bra{NN;\,\vec{p},\alpha}}P_{\Delta}
 \Big(
 {\cal J}^{[2]\,\mu}_{\Delta{\bar N}}(\vec{k}) +
 {\cal J}^{(\pi)/r \, \mu}_{\Delta}(z,\vec{k}\,)  {\hat R}^{\pi}(z-k) \Big)
 P_{\bar N} \ket{d'}\,, 
  \label{kap5_eff_matrix_dd2}
\end{eqnarray} 
where the quantities   ${\cal J}^{[2]\,\mu}_{\Delta{\bar N}}(\vec{k})$ and 
${\cal J}^{(\pi)/r \, \mu}_{\Delta}(z,\vec{k}\,)$ are defined in analogy to 
(\ref{kap5_j_n2_ren_2}).  

\subsection{The effective current operator in the static limit}
\label{kap5_statisch_current}

Until now, we have only considered the {\it retarded} form of the effective
current. In the static case, considerable simplifications occur. Whereas
the one-body contributions remain unaltered (apart from the dressing factors 
which are, of course, not present in static calculations because in this case 
only physical instead of bare nucleons appear), we have to substitute the
retarded meson-$NN$ propagators in Eqs.~(\ref{kap5_kontakt_1}) through 
(\ref{kap5_diss_2}) by their static limits, i.e.\ 
\begin{equation}\label{kap5_stat_prop}
 G_0(z)\mbox{ and } G_0(z-k) 
\longrightarrow -\frac{1}{h_x} \,\, ,
\end{equation}
where $h_x$ denotes the kinetic energy of a meson $x$. Consequently, 
the pole structure of the MEC is lost. The meson-nucleon 
vertices in the corresponding expressions are  those used in the
static Bonn-OBEPR potential \cite{MaH87} so that the sum of the 
pion-in-flight, vertex and contact MEC fulfils current conservation with the
$\pi$ exchange part $V^{\pi}_{OBEPR}$ of the Bonn-OBEPR potential
\begin{eqnarray}
 \vec{k} \cdot \left(
 \vec{J}_{stat}^{\,\pi/c}(\vec{k}) + \vec{J}_{stat}^{\,\pi/f}(\vec{k})
  + \vec{J}_{stat}^{\,\pi/v}(\vec{k}) \right)
  &=&  \left [
 V^{\pi}_{OBEPR}\,, \, {\rho}_{real}^{nr}(\vec{k}) \right] \, . 
 \label{kap5_kontin_pi_mec_1_stat}
\end{eqnarray}
Note however, that a static  recoil contribution does not appear, 
because in the static limit the wave function renormalization current 
cancels exactly the recoil current \cite{GaH76,Are93}. Likewise, the recoil 
contribution in the  $\Delta$ MEC (Fig.~\ref{figem15}) has to be neglected, 
too.  

\subsection{Inconsistencies in static approaches}\label{kap5_stat_inkon}

\begin{figure}[tp]
\centerline{\psfig{figure=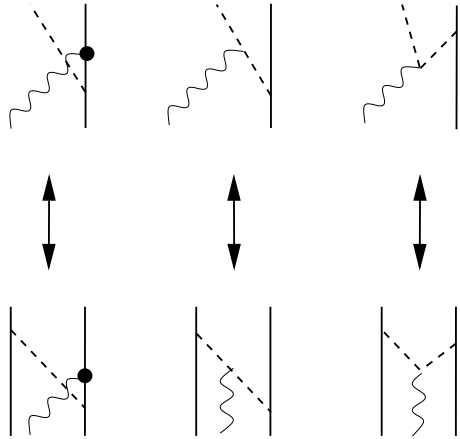,width=6cm,angle=0}}
\vspace{0.5cm}
\caption{Correspondence of Born terms of pion photoproduction to $\pi$ MEC.}
\label{figem17}
\end{figure}

At the end of the discussion of the effective current operator, we would 
like to point out some inconsistencies in static approaches, namely the 
fact that the $\Delta$ current and the $\pi$ MEC are {\it not} 
independent from each other. From Fig.~\ref{figem17} it becomes evident that
the Born terms of the $M_{1+}^{(3/2)}$ amplitude are related to a part of the 
$\pi$ MEC in the two-nucleon system, namely to a part of the pion-in-flight 
and the recoil current. A static treatment of the MEC leads therefore to 
serious inconsistencies. For example, the crossed nucleon pole graph is not 
contained in the static $\pi$ MEC because the respective recoil MEC is not 
present at all in static approaches due to its cancellation against the 
wave function renormalization current. How critical this inconsistency is 
has been demonstrated by Wilhelm {\it et al.}~\cite{Wil92,WiA93} in determining 
another effective $\gamma {\bar N} \Delta$ coupling, denoted henceforth by 
$\widetilde{G}_{M1}^{\Delta {\bar N}}(\mbox{eff3})$, in a fit to the data 
where the Born amplitude in (\ref{kap5_gamma_pi_2_1}) is set equal to zero 
so that it is contained effectively in this coupling. 
At least below the resonance position this ``modified'' coupling is 
considerably larger as the original coupling (compare the dashed and  
dotted curves in Fig.~\ref{figem10}). In view of 
this rather large difference between the two couplings, 
Wilhelm {\it et al.}~\cite{WiA93}
 had already suspected that retardation in the 
MEC may become important in the $\Delta$ region. Using the modified 
parameters, a much improved description of the total cross section was 
achieved (see Sect.~\ref{kap6_deutspalt}), though the 
differential cross section was still poorly described in the $\Delta$ region.
Moreover, it is obvious that the modified coupling is conceptually not 
very satisfying because the neglect of the Born amplitude leads formally 
to a vanishing rescattering 
amplitude, so that the resulting effective $\gamma {\bar N} \Delta$ coupling
is identical to the bare one and therefore energy independent, in contrast
to the present parametrization in Eq.~(\ref{kap5_parametrisierung_1}).

\section{The question of gauge invariance}\label{kap5_eichinvarianz_2}

Now, we would like to discuss the question whether the effective current 
$J_{eff}^{\mu}(z,\vec{k})$ is gauge invariant, i.e., is consistent with 
the hadronic interaction. As first, we will formulate the continuity equation
for the {\it effective} operators. From the continuity equation for the 
{\it full} current operator
\begin{equation}\label{kap5_kontin_1_wieder}
 \bra{f} \left(  \left[ H,J^0(\vec{k}) \right] \,  -   \, 
   \vec{k} \cdot  \vec{J}(\vec{k}) \right) \ket{i} =0\, 
 \end{equation}
one obtains straightforwardly the relation
\begin{equation}\label{kap5_kontin_eff}
 \bra{f}P \,  \vec{k} \cdot \vec{J}_{eff}(z_f,\vec{k})\, P \ket{i}
  = \bra{f}P  \left( H_{eff}(z_f)  \rho_{eff}(z_f,\vec{k})
  - \rho_{eff}(z_f,\vec{k}) H_{eff}(z_i) \right)P \ket{i}\,, 
\end{equation}
where $z_{i/f}=E_{i/f}\pm i\epsilon$, and the interaction part $V_{eff}(z)$ of 
the effective Hamiltonian consists of a disconnected and a connected part 
\begin{eqnarray} 
 V_{eff}(z) &=& V^{dis}_{eff}(z) + V^{con}_{eff}(z)  \,,
 \label{kap5_v_eff_2}
 \end{eqnarray}
 with
 \begin{eqnarray}
 V^{dis}_{eff}(z) &=& V^{[c]}_{PP} 
 + \left[V^0_{PX} G_0(z) V^0_{XP}\right]_{dis}\, ,\label{kap5_v_eff_3}\\
 V^{con}_{eff}(z) &=&  V_{PP}^{0\, [2]} + 
 \left[V^0_{PX} G_0(z) V^0_{XP}\right]_{con}
 \, .\label{kap5_v_eff_4}
\end{eqnarray}
Note that we have neglected $V^0_{XX}$ in $V^{con}_{eff}$ in view of the 
discussion in Sect.~\ref{kap5_effektiv_n2}. 

Splitting the hadronic interaction and the current into one- and two-body 
parts, one obtains 
\begin{eqnarray}
 \bra{f} P \,\vec{k} \cdot \vec{J}^{\,[1]}_{eff}(z_f,\vec{k}) P \,\ket{i}
  &=& \bra{f} P \left[ \left(H_{0\,PP}+ V^{dis}_{eff}(z_f) \right)
   \rho^{[1]}_{eff}(z_f,\vec{k}) 
 - \rho^{[1]}_{eff}(z_f,\vec{k}) \left( H_{0\,PP} + 
 V^{dis}_{eff}(z_i) \right) \right] 
  P \,\ket{i}\, , \label{kap5_effektiv_kontin_diss_1}\\
 \bra{f} P\, \vec{k} \cdot \vec{J}^{\,[2]}_{eff}(z_f,\vec{k}) P \,\ket{i}
  &=& \bra{f} P \Big[ V^{con}_{eff}(z_f) 
  \Big( \rho^{[1]}_{eff}(z_f,\vec{k}) + \rho^{[2]}_{eff}(z_f,\vec{k})\Big)  
\nonumber\\&&
- \Big( \rho^{[1]}_{eff}(z_f,\vec{k}) +\rho^{[2]}_{eff}(z_f,\vec{k}) \Big)
 V^{con}_{eff}(z_i)  \Big] P \ket{i}\, . \label{kap5_effektiv_kontin_con_1}
\end{eqnarray}
These equations serve as starting point for the discussion of gauge 
invariance. We will restrict ourselves to the pure nucleonic currents and 
interactions acting solely in ${\cal H}_{{\bar N}}$ for the sake of 
simplicity. Concering the one-body part in 
(\ref{kap5_effektiv_kontin_diss_1}), it is clear 
from the very beginning that in retarded calculations, where $V^{dis}_{eff}$ 
is not zero, gauge invariance cannot exactly be fulfilled. Note that we 
do not evaluate explicitly the loops in Fig.~\ref{figem4}
which enter into $J_{eff}^{N[1]\, \mu}(z,\vec{k})$ as has been discussed 
in subsection \ref{kap5_jnn}. On the other hand, the corresponding loop
in $V^{dis}_{eff}$ has to be evaluated due to the requirement of unitarity
 of the hadronic interaction. 
Therefore, Eq.~(\ref{kap5_effektiv_kontin_diss_1}) is only valid in the 
onshell case, i.e., for
\begin{equation}\label{kap5_onshell_1}
 P_{\bar N} \ket{i} = 
 \ket{{\bar N}{\bar N};\,\vec{p}_i, \alpha_i}\, ,
  \quad P_{\bar N}\ket{f} =
 \ket{{\bar N}{\bar N};\,\vec{p}_f, \alpha_f}\, ,
\end{equation}
where $\vec{p}_{i/f}$ is given by $E_{i/f}\, = \, 2 e_N^{nr}(\vec{p}_{i/f})$. 
Note that in this case $V^{dis}_{eff}$ is exactly zero. But 
in static approaches, Eq.~(\ref{kap5_effektiv_kontin_diss_1}) can always be 
fulfilled because of $V^{dis}_{eff} \equiv 0$.

\begin{figure}[ttt]
\vspace{0.5cm}
\centerline{\psfig{figure=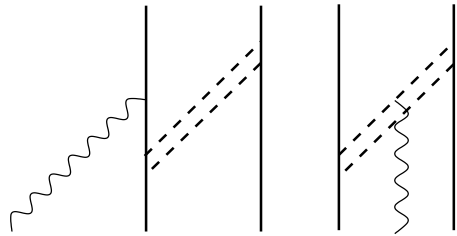,width=6cm,angle=0}}
\vspace{0.5cm}
\caption{Diagrammatic representation of retarded MECs
 which are of fourth order in the pion-nucleon coupling constant.}
 \label{figem16}
\end{figure}

We now turn to the discussion of the two-body part in
(\ref{kap5_effektiv_kontin_con_1}). According to 
(\ref{kap5_effektiv_n2_expl_1}), the effective current operator 
$J_{eff}^{N[2]\, \mu}(z,\vec{k})$ consists of the $\pi$ and $\rho$ MEC and 
the $\gamma\pi\rho /\omega$ currents. With respect to the 
$\rho$ MEC, we refer to the discussion at the end of subsection \ref{kap5_jnn}.
Concerning the $\gamma\pi\rho /\omega$ contribution, one should 
notice that the corresponding charge operator vanishes due to the use of 
nonrelativistic vertex structures, whereas the remaining spatial part 
${\vec J}^{\, (diss)}(z,\vec{k}\,)$ is purely transverse
\begin{equation}\label{kap5_gpr_kont}
 \vec{k} \cdot {\vec J}^{\, (diss)}(z,\vec{k}\,) =0\, .
\end{equation}
Consequently, (\ref{kap5_effektiv_kontin_con_1}) is fulfilled with respect to
the dissociation current. The remaining retarded $\pi$ MEC satisfies 
($z_i = z_f -k$)
\begin{eqnarray}
 \vec{k} \cdot \vec{J}^{\,(\pi)}(z_f,\vec{k})
  &=&  V^{\pi\,con}_{eff}(z_f) {\rho}_{real}^{nr}(\vec{k}) -
  {\rho}_{real}^{nr}(\vec{k}) V^{\pi\,con}_{eff}(z_i) +
  k J^{(\pi)\, 0}(z_f,\vec{k})   + {\cal A}(z_f, \vec{k}\,)
  + {\hat {\cal O}}\left(\frac{1}{M_{N}^3}\right)
 \label{kap5_kontin_pi_mec_1}
\end{eqnarray}
with the retarded $\pi$ exchange potential 
\begin{equation}\label{kap5_kontin_pi_pot_1}
 V^{\pi\,con}_{eff}(z) = 
 \left[ V^{0\, nonrel}_{{\bar N} X_{\pi}}\, G_0(z) 
 V^{0\, nonrel}_{X_{\pi}{\bar N}}\, \right]_{con}\,
\end{equation}
-- note that in (\ref{kap5_kontin_pi_pot_1}) 
$V^{0\, nonrel}_{{\bar N} X_{\pi}}$ occurs due to the use of the 
nonrelativistic pion-nucleon vertex in the MEC --, and the auxiliary quantity
 \begin{eqnarray}
 {\cal A}(z, \vec{k}\,) &=&
  \sum_{i,j=1,2 ;\,j \neq i}\, 
 \left\{ \left(
  v^{0\, nonrel}_{{\bar N} X_{\pi}}(i)
  G_0(z) v^0_{X_{\pi} X_{\pi \pi}}(j)  G_0(z)
  \, {\rho}^{(0)}_{X_{\pi \pi} {\bar N} }(\vec{k}) \right)
  \, \left(E_i - H_{0\,{\bar N}{\bar N}} \right) \right.\nonumber\\
  & & \left. \quad -
 \left(E_f - H_{0\, {\bar N}{\bar N}} \right) \,
  \left(
   \, {\rho}^{(0)}_{{\bar N} X_{\pi \pi}} (\vec{k})\,
 G_0(z-k)
  v^{0\, nonrel}_{X_{\pi \pi}X_{\pi}}(i) G_0(z-k) v^{0}_{X_{\pi}
  {\bar N}}(j) \right) \right\}\, . 
 \label{kap5_a}
\end{eqnarray}
In general, only between plane waves  the matrixelement of 
${\cal A}(z, \vec{k}\,)$ vanishes, which fact 
indicates that in a retarded approach additional MECs of at least
fourth order in the pion-nucleon coupling are necessary for ensuring
gauge invariance. Examples for such contributions are depicted in 
Fig.~\ref{figem16}. They are quite complicated and beyond the 
scope of the present work.
 In the static limit such MECs are not present because 
 they either cancel out  each other  or cancel 
 against corresponding wave function renormalization currents 
so that (\ref{kap5_effektiv_kontin_con_1}) is satisfied. 

In summary, gauge invariance is fulfilled 
up to second order in the meson-nucleon coupling constant, whereas violations 
appear in higher order. However, we hope that by the use of the Siegert 
decomposition of the electric multipoles~\cite{ArS91} at least a part of these 
neglected higher order contributions are taken into account implicitly.

\section{Results for deuteron photodisintegration}\label{kap6_deutspalt}

We now will discuss the results for deuteron photodisintegration. The 
formalism for calculating the various observables may be found in the 
review~\cite{ArS91}. As currents we consider all currents discussed before, 
i.e., the one-body spin and convection current as well as
the spin-orbit current and $\pi$ and $\rho$ MEC. The parameters of 
the coupling constants are chosen consistently with the underlying 
$NN$ interction. We have included all multipoles up to $L \leq 4$.  
For the electric transitions we have used the Siegert operators~\cite{ArS91} 
which incorporate a large fraction of MEC contributions. They include 
besides the usual one-body density also two-body charge terms as described 
in  Sect.~\ref{kap5_effektiv_n2}. The remaining MEC contributions beyond 
the Siegert operators are evaluated explicitly. In view of the fact, that 
retardation is most important at short distances, we have considered 
retardation in the  contact, vertex,  meson-in-flight and dissociation 
MECs only for multipoles up to $L \leq 2$, while 
for $L > 2$ the static expressions are used, which is sufficient as we have 
checked in detail. 

\subsection{Retardation effects in a pure nucleonic model}
\label{kap6_vergleich_obepr_elster}

We will begin the discussion by studying first the influence of retardation 
in a pure nucleonic model, where the $\Delta$ d.o.f.\ as well as the 
$\pi d$ channel are switched off so that only a pure $NN$ interaction and 
the corresponding currents are taken into account. For the sake of simplicity, 
the dissociation current is also switched off. In detail, we consider four 
different types of calculations which differ with respect to static or 
retarded potentials and MECs, comprising contact, vertex and 
meson-in-flight currents, and with respect to the recoil MEC. We 
will denote them by N(potential, current, recoil current), and list the four
cases which we consider in Table~\ref{kap6_tab_3}. As static potential we use
the Bonn OBEPR model, while for the retarded interaction we take the Elster
model. 

\begin{figure}[hhp]
\centerline{\psfig{figure=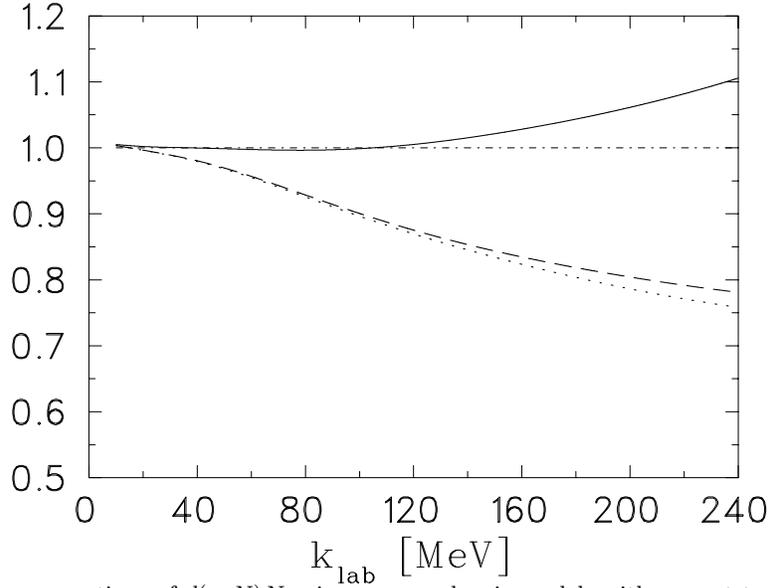,width=10cm,angle=90}}
\caption{Ratios of total cross sections of $d(\gamma,N)N$ using pure 
nucleonic models with 
respect to $\sigma_{tot}(\mbox{N(stat,stat,0)})$:  dotted curve:  
N(ret,stat,0), dashed curve: N(ret,ret,0),  full curve: N(ret,ret,1).}
\label{figem18}
\end{figure}

It is obvious that due to the neglect of explicit $\Delta$ d.o.f. no 
realistic description is possible above about $k_{lab} = 100$ MeV. 
But here we are interested mainly in the relative sensitivity of the cross 
sections with respect to the various static and retarded approaches. For 
this reason we show in Fig.~\ref{figem18} only the 
ratios of the total cross sections with respect to N(stat,stat,0). 
One readily notices that retardation in the hadronic interaction, i.e., 
going from N(stat,stat,0) to N(ret,stat,0), leads to a drastic reduction
of the cross section which amounts to about 25 percent at $k_{lab}$ = 240 MeV.
On the other  hand, retardation in the meson-in-flight, contact and vertex 
MECs is almost negligible. It turns 
out that retardation is important only in the lowest multipole transitions to 
the $^1S_0$ and $^3P_0$ partial waves, which are of minor importance  
in the total cross section, except directly at threshold. In general, the 
influence of retardation in the MEC decreases rapidly with increasing 
multipole order. On the other hand, the recoil current contribution turns 
out to be very important as is evident comparing N(ret,ret,0) with 
N(ret,ret,1) in Fig.~\ref{figem18}. It is of particular 
importance in the contributions of the  $^1D_2$ and $^3F_2$ partial waves.

In summary, it turns out that below $\pi$ threshold, the static and the
full retarded calculations predict the same cross section within a few percent.
This result is at variance with the previous work of Schmitt 
{\it et al}.~\cite{ScA89b}, in which retardation in potential and current
had been considered using a $p/M_N$ expansion below $\pi$ threshold. 
A sizeable decrease of the cross section was found in retarded calculations  
compared to a static approach. In contrast to the present work, 
also the wave function renormalization current was included but no 
renormalization of the continuum wave functions was applied. Only the bound 
deuteron wave function had been renormalized globally by a constant. 
In a correct approach, however, the wave function renormalization current
should have been left out as well as the renormalization of the deuteron state. 
This erroneous treatment lead to the mentioned decrease of the cross section. 

In the energy region above $\pi$ threshold, the equivalence between 
static and retarded frameworks breaks down, the difference increasing 
with increasing energy as can be 
seen in Fig.~\ref{figem18}. This is not surprising, 
because above $\pi$ threshold the pion can become an onshell particle 
which fact is only taken into account by the retarded $\pi NN$ propagators, 
whereas for a static framework no coupling between the $NN$ and the open 
$\pi NN$ channel is possible, i.e., retarded and static approaches  
exhibit a different behaviour.

\subsection{Retardation effects in the $\Delta$ resonance region}
\label{kap6_vergleich_voll} 

Now we will turn to the energy region of the $\Delta$ resonance, applying our
full model which includes explicitly $\Delta$ degrees of freedom. 
In detail, we will consider four different static and five different retarded
approaches whose nomenclature is explained in Tables~\ref{kap6_tab_stat} 
and \ref{kap6_tab_ret}. The corresponding hadronic interaction models
 are described in detail in I and are therefore not repeated in this paper. 
We have restricted the full inclusion of $\Delta$ 
d.o.f.\ to partial waves with total angular momentum $j \leq 3$, and use 
for $j \geq 4$ -- similar to \cite{WiA93} -- the impulse 
approximation~\cite{ArD71} in which the $N\Delta$ components are given  
in first order perturbation theory by
\begin{eqnarray}
 P_{\Delta} \ket{NN;\,\vec{p}, \alpha}^{(-)} 
&=& G_0^{(\Delta)}(z)\,V^{con}_{[2]\,\Delta{\bar N}}(z)
 P_{\bar N} \ket{NN;\,\vec{p}, \alpha}^{(-)}\,,
\label{kap6_stoss2}
\end{eqnarray}
where the nucleonic component $\ket{NN;\,\vec{p}, \alpha}^{(-)}$ is obtained 
in pure nucleonic space from a realistic $NN$ potential.

\begin{figure}[b!]
\centerline{\psfig{figure=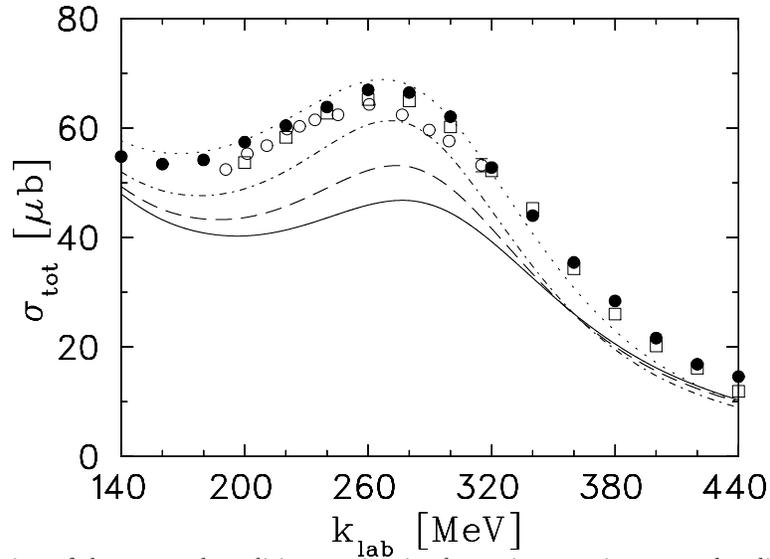,width=10cm,angle=90}}
\caption{Total cross section of deuteron photodisintegration in the various
static approaches listed in Table~\ref{kap6_tab_stat}: full curve CC(stat1),
dashed curve CC(stat2), dash-dotted curve CC(stat3), and dotted curve 
CC(stat4). Experimental data from {\protect \cite{CrA96}} ($\bullet$), 
 {\protect \cite{ArG84}} ($\Box$) and
{\protect \cite{BlB95}} ($\circ$). 
}
\label{figem19}
\end{figure}

\begin{figure}[p!]
\centerline{\psfig{figure=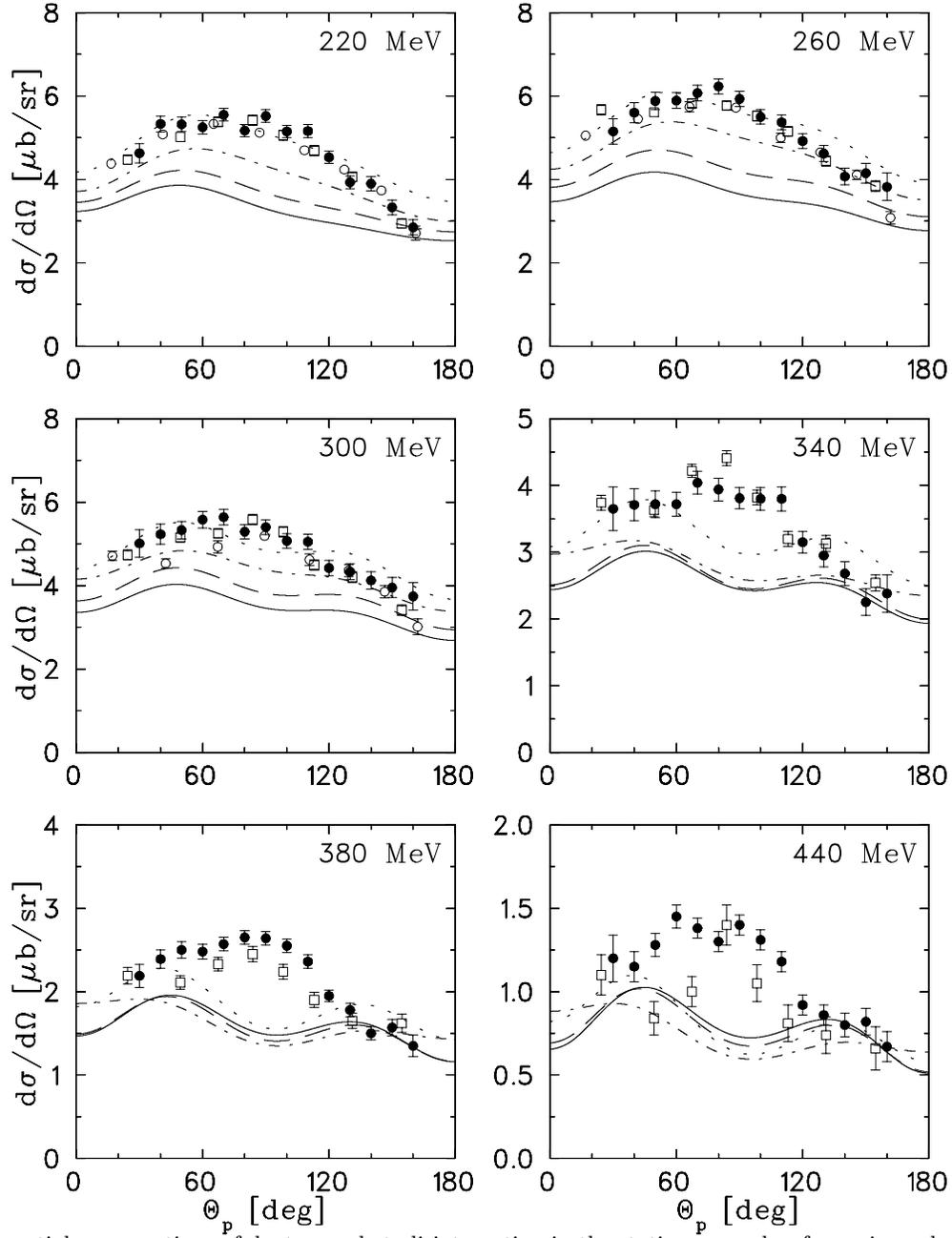,width=13cm,angle=0}}
\caption{Differential cross sections of deuteron photodisintegration in the 
static approaches for various photon energies $k_{lab}$ as function of 
the c.m.\ proton angle $\theta_p$. Notation of the curves and experimental 
data as in Fig.~\ref{figem19}.}
\label{figem20}
\end{figure}

As first we consider the results of the static MEC cases in 
Figs.~\ref{figem19} and \ref{figem20}. The
model CC(stat1), which has already been used by Wilhelm 
{\it et al}.~\cite{WiA93}, shows the already mentioned significant 
underestimation of the total cross section by about 30~\% at 
$k_{lab} = 260$~MeV. Moreover, above $k_{lab} =300$~MeV the differential cross 
sections develop a dip structure around $90^\circ$ 
in clear contradiction to the experimental 
data. Due to the slightly stronger $\gamma {\bar N} \Delta$ coupling in 
the model CC(stat2) -- compare the couplings 
${\widetilde G}_{M1}^{\Delta {\bar N}}(\mbox{eff}\,1)$ and
${\widetilde G}_{M1}^{\Delta {\bar N}}(\mbox{eff}\,2)$ in 
Fig.~\ref{figem10} -- this model leads
to a noticeable enhancement of $\sigma_{tot}$ which is, however, not
sufficient to bridge the gap to experiment. The additional improvements, 
in the approach CC(stat3), especially the incorporation of the 
$\gamma\pi\rho/\omega$ currents, lead to a further enhancement 
of the cross section for $k_{lab} 
\stackrel{\textstyle <}{_{_{\textstyle \widetilde{ }}}} 340$ MeV.
Compared to the recent Mainz data \cite{CrA96}, the theoretical
underestimation of $\sigma_{tot}$ amounts to not more than 10~\%
at $k_{lab}=  260$~MeV but increases to about 40~\% at $k_{lab}=440$ MeV. 
Moreover, the dip problem in the differential cross section has not 
been resolved. The last static model CC(stat4) shows, as has already been
discussed in \cite{WiA93}, a rather good description of the total
cross section in the peak region due to the substantially stronger 
$\gamma {\bar N} \Delta$ coupling, but again the dip problem as well as
the underestimation of the cross section at higher energies remains.
Moreover, as has been stressed in Sect.~\ref{kap5_stat_inkon},
this approach is quite unsatisfactory from a conceptual point of view.

\begin{figure}[b!]
\centerline{\psfig{figure=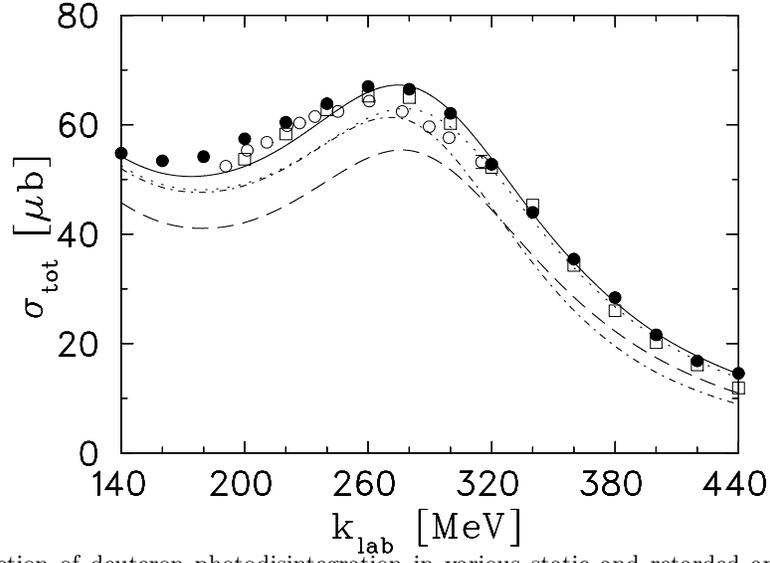,width=10cm,angle=90}}
\caption{Total cross section of deuteron photodisintegration in various 
static and retarded approaches. Notation of the curves (see 
Tables~\ref{kap6_tab_stat} and \ref{kap6_tab_ret} for the nomenclature): 
dash-dotted CC(stat3), dashed curve CC(ret4), full curve CC(ret2), 
dotted curve CC(ret5). Experimental data as in Fig.~\ref{figem19}.}
\label{figem21}
\end{figure}

\begin{figure}[p]
\centerline{\psfig{figure=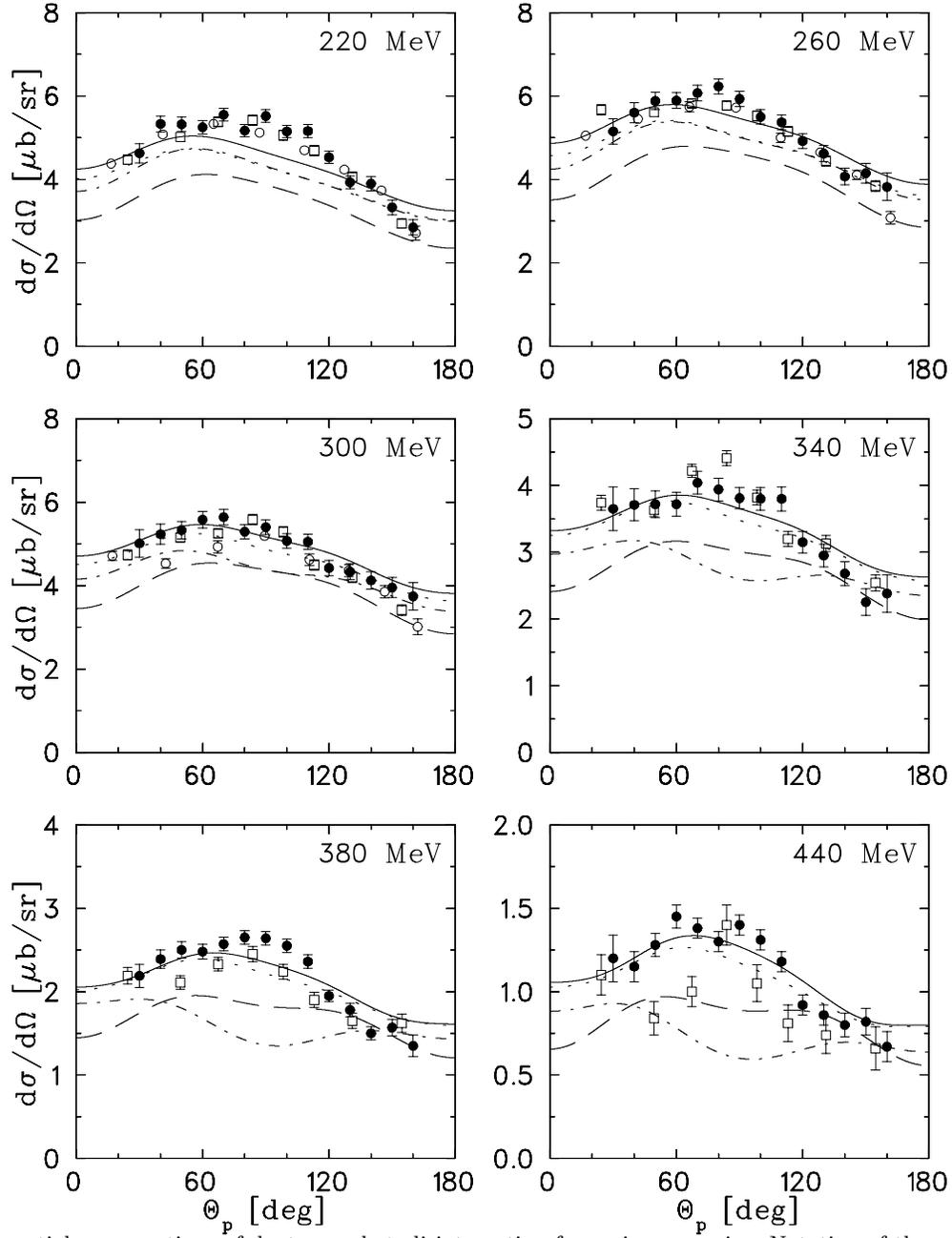,width=13cm,angle=0}}
\caption{Differential cross sections of deuteron photodisintegration for 
various energies. Notation of the curves and experimental data as in 
Fig.~\ref{figem21}.}
\label{figem22}
\end{figure}

The results for the retarded current models are shown in 
Figs.~\ref{figem21} and \ref{figem22}. 
The effect of introducing retardation in the hadronic interaction but 
keeping the currents still static can be seen by comparing CC(stat3)  
with CC(ret4). One readily notes a drastic decrease of the cross section.
This fact has been noted already for the pure nucleonic case discussed in 
the previous subsection. Switching on in addition retardation in the currents
(transition CC(ret4) $\rightarrow$ CC(ret2)) leads, however, to a drastic 
increase which overcompensates the reduction of the cross section by 
the retardation in the hadronic interaction. It turns out that within about 
10~\% an overall satisfactory description of the total cross section
over the whole energy region is achieved. Concerning the differential
cross sections, apart from energies below the resonance maximum, the
description of the experimental data is almost quantitative, too. Especially 
the above mentioned dip structure occurring in the static treatment vanishes 
{\it completely}. This is mainly due to the additional consideration of the 
$N$-$\Delta$-mass difference in  $V^0_{N \Delta}$ and $V^0_{\Delta \Delta}$,
and due to the recoil current contributions. On the other hand, similar to 
the pure nucleonic case, retardation in the pion-in-flight, vertex, contact 
as well as in the dissociation currents is negligible in the total and 
differential cross sections. With respect to the influence of the 
$\pi d$ channel, it turns out that this mechanism  leads to a noticeable 
enhancement below about $k_{lab}$ = 300 MeV. This result is at variance 
with the results of Tanabe and Ohta~\cite{TaO89} who found a 
decrease of $\sigma_{tot}$ when they incorporated the $\pi d$ channel. 
On the other hand, the influence of the $\pi d$ channel on the 
$^1D_2$ channel is qualitatively the same than in our approach. Thus,
the reason for this difference is unclear.   

\begin{figure}[p]
\centerline{\psfig{figure=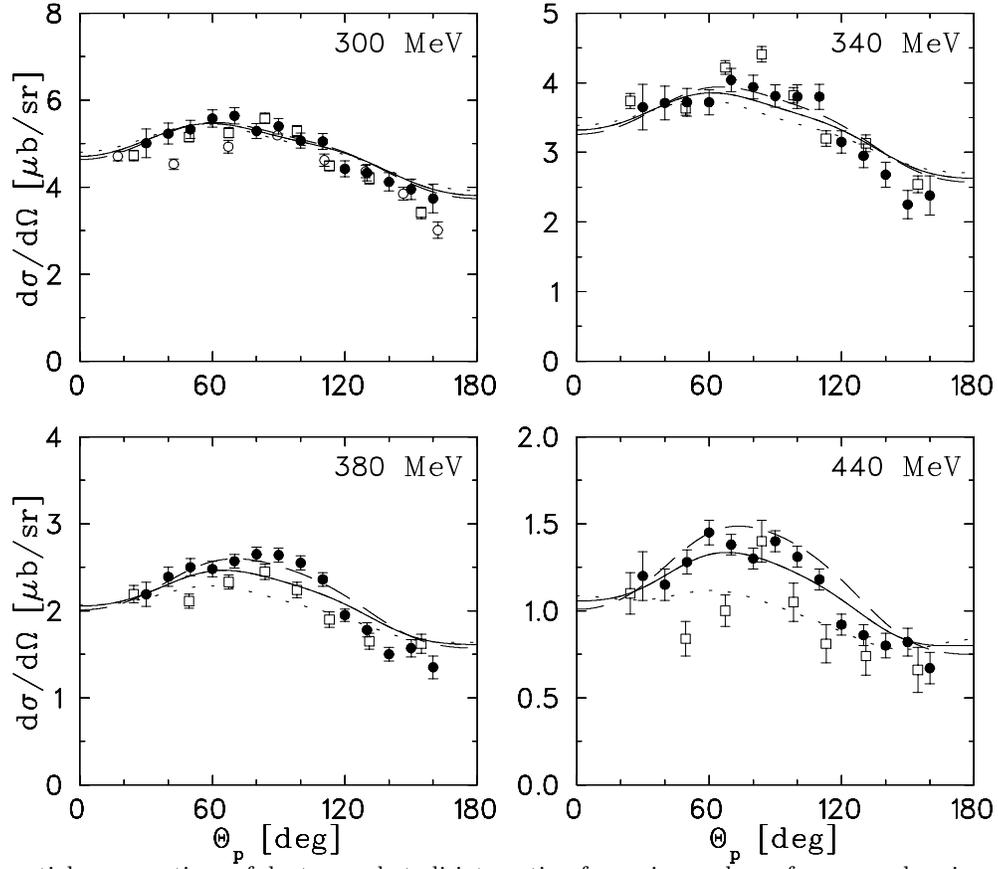,width=13cm,angle=0}}
\caption{Differential cross sections of deuteron photodisintegration for 
various values of $\alpha_{\Delta N \rho}$ and various photon energies.
  Notation of the curves:
  dashed curve CC(ret1), full curve CC(ret2), dotted curve CC(ret3).  
  Notation of the  experimental data as in 
Fig.~\ref{figem19}.}
\label{figem30}
\end{figure}

As next topic, we discuss in Fig.~\ref{figem30}  
 the influence of the parameter 
$\alpha_{\Delta N \rho}$ describing the additional $\rho N \Delta$ coupling 
in Eq.~(76) of I. Comparing the results of CC(ret1), CC(ret2) and CC(ret3), 
which differ by the value of $\alpha_{\Delta N \rho}$, but which are 
equivalent concerning the description of the $^1D_2$ channel in $NN$ 
scattering, it turns out that the value of $\alpha_{\Delta N \rho}$ is of 
considerable importance for energies above about $k_{lab}= 380$ MeV. But 
similar to $NN$ scattering, an optimal value for $\alpha_{\Delta N \rho}$
could not be determined because of the rather large differences between the 
different experimental data sets at higher energies~\cite{CrA96,ArG84}. 

\begin{figure}[bp!]
\centerline{\psfig{figure=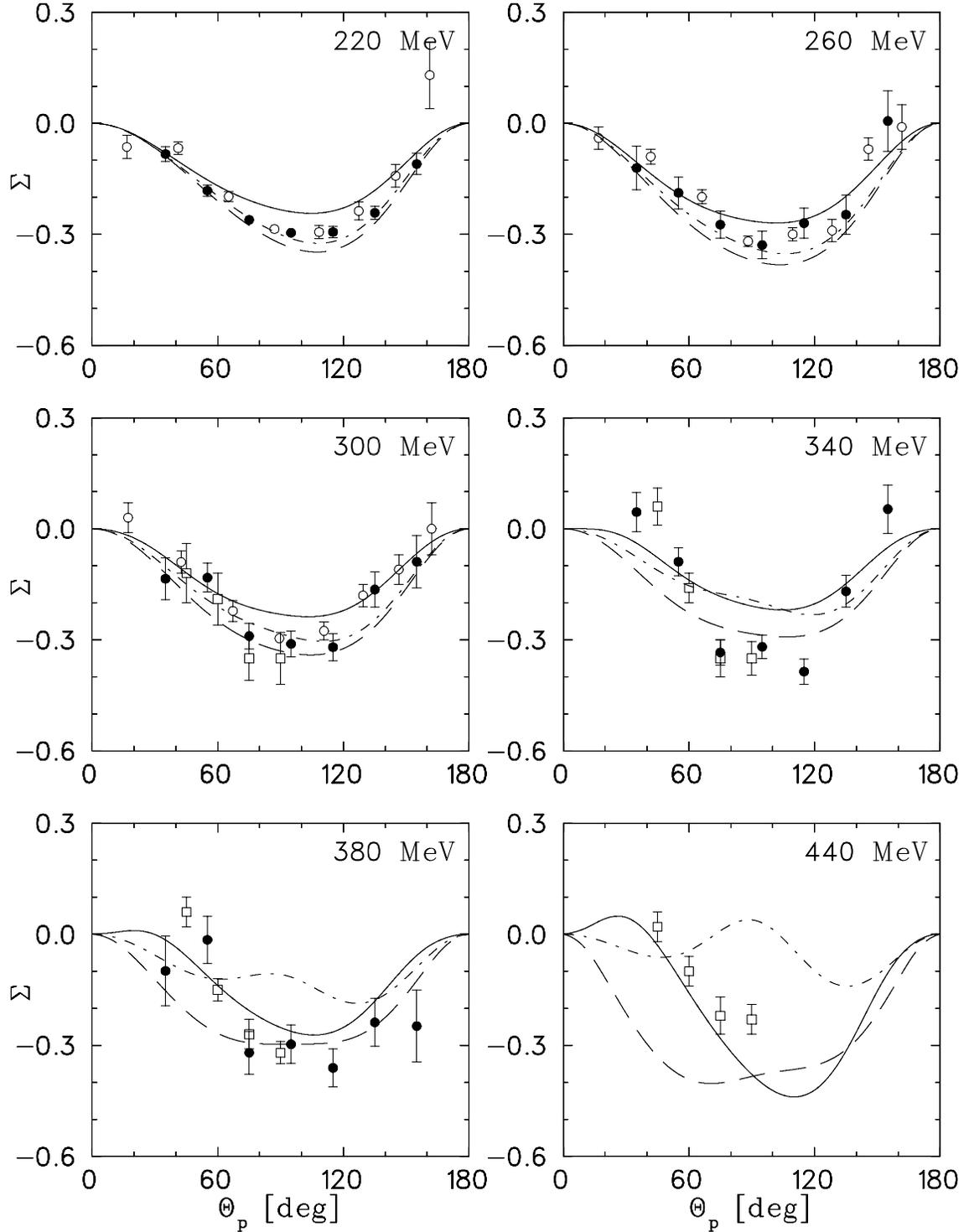,width=15cm,angle=0}}
\caption{Linear photon asymmetry $\Sigma$ of deuteron photodisintegration. 
Notation of the curves: dash-dotted CC(stat3), dashed CC(ret4), 
full CC(ret2). Experimental data from 
{\protect  \cite{BlB95}} ($\circ$), 
{\protect  \cite{WaA98}} ($\bullet$) and
{\protect  \cite{AdB91}} ($\Box$).
 } 
\label{figem23}
\end{figure}

Finally, we discuss a few polarization observables. At first, we will
 consider   the 
linear photon asymmetry $\Sigma$ for which new experimental data is 
available~\cite{BlB95,WaA98}. In Fig.~\ref{figem23}, the 
results for the approaches CC(stat3), CC(ret4) and CC(ret2) are depicted. 
The static approach CC(stat3) yields quite a good description for energies 
up to about $k_{lab}=300$ MeV. For higher energies, however, an  
oscillatory structure appears which  vanishes completely
if retardation in the hadronic interaction is switched on 
(CC(stat3) $\rightarrow$ CC(ret4)). Moreover, $\Sigma$ decreases drastically
with increasing photon energy. The additional inclusion of retardation 
in the currents (CC(ret4) $\rightarrow$ CC(ret2)) leads to an 
overall reduction of $\Sigma$ except for the highest energy for which a
significant change of the shape appears. 
Compared to the experimental data one finds
a satisfactory overall description of the qualitative features.
But one notes also some significant deviations around $90^{\circ}$,
where theory predicts an asymmetry considerably smaller than experiment. 
This is particularly significant at $k_{lab}= 220$ MeV in view of the small
error bars, while for higher energies the larger errors make the deviations
less significant. It is interesting to note that for $k_{lab}= 220$ MeV
also the differential cross section exhibits around $90^{\circ}$ a sizeable
deviation of the theory by about 10~\%. 

\begin{figure}[bp!]
\centerline{\psfig{figure=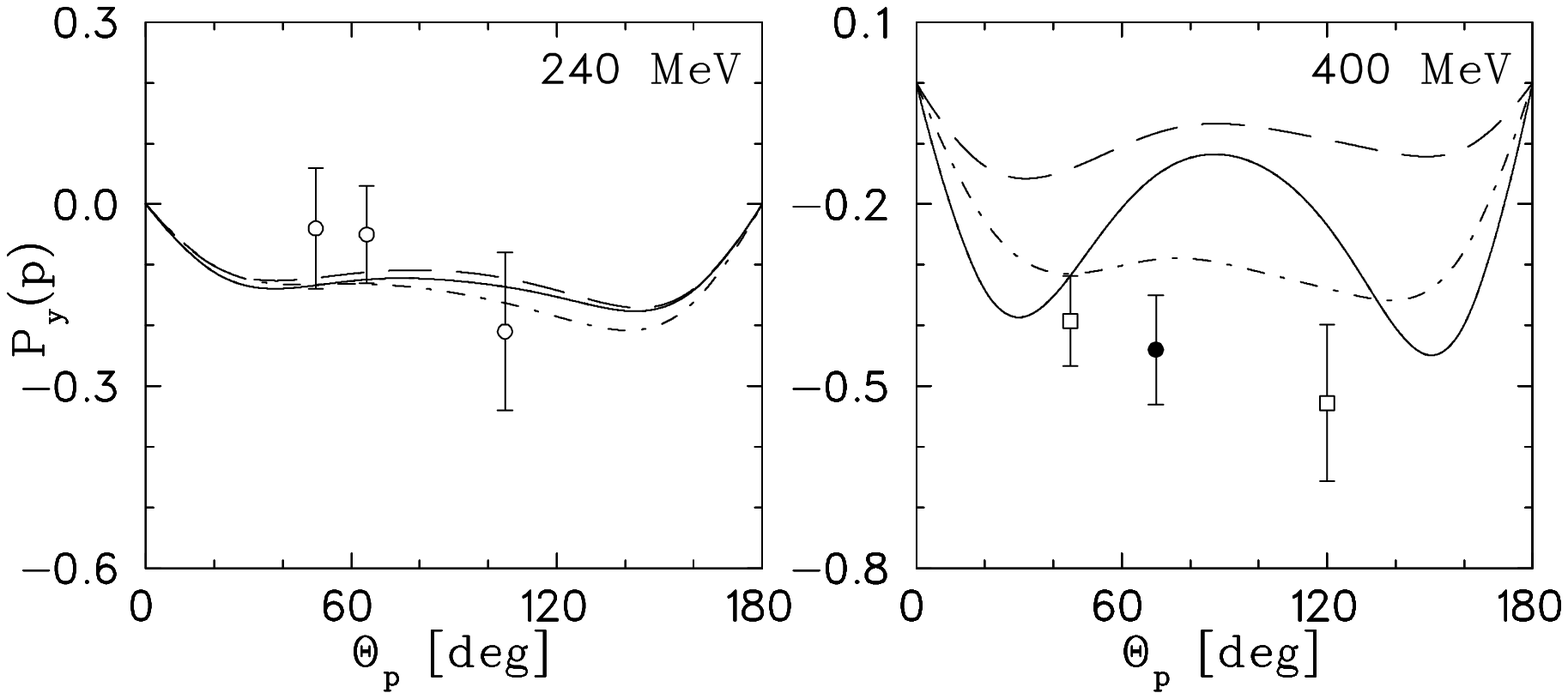,width=15cm,angle=0}}
\caption{Proton Polarization $P_y(p)$ of deuteron photodisintegration. 
Notation of the curves: dash-dotted CC(stat3), dashed CC(ret4), 
full CC(ret2). Experimental data from {\protect  \cite{BrD80}} ($\Box$), 
{\protect  \cite{LiL68}} ($\circ$), and {\protect  \cite{IkA79}} ($\bullet$).}
\label{figem24}
\end{figure}

\begin{figure}[bp!]
\centerline{\psfig{figure=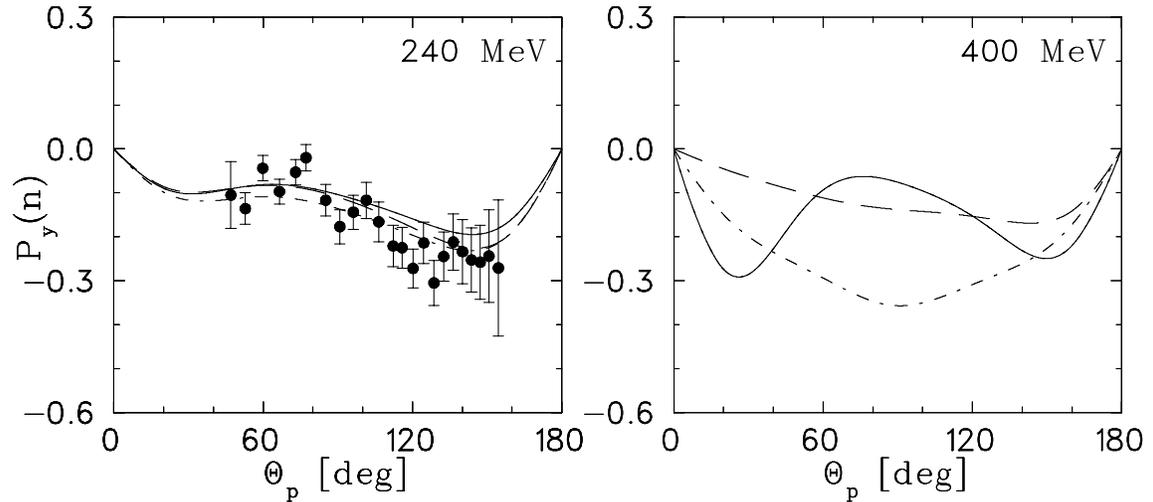,width=15cm,angle=0}}
\caption{Neutron Polarization $P_y(n)$ of deuteron photodisintegration. 
Notation of the curves: dash-dotted CC(stat3), dashed CC(ret4), 
full CC(ret2). Experimental data from {\protect  \cite{HuC87}} ($\bullet$).}
\label{figem25}
\end{figure}

Finally, we show in Figs.~\ref{figem24} and \ref{figem25}
the proton and neutron polarization, respectively. At $k_{lab}=240$ MeV, 
the different approaches yield very similar
results which are also in good agreement with the experimental data.
At $k_{lab}=400$ MeV hadronic retardation results in a strong
overall decrease of the polarization for both neutron and proton
(transition CC(stat3) $\rightarrow$ CC(ret4)). Switching
on the current retardation leads however to a strong polarization increase 
in the forward and backward direction while around $90^{\circ}$  the 
polarization is much less affected. This feature is clearly at variance 
with the few proton 
polarization  data. Due to the fact that in general polarization 
observables are more sensitive to small reaction mechanisms, one reason 
for this failure may be the fact that apart from the dominant 
$^1D_2$ channel several other partial waves (for example the $P$- and the
$^3F_3$ waves) are described only fairly well within the present model 
(see I for further details). Thus, one has to await future improvements of the
hadronic interaction model in order to see whether this failure could be 
resolved. Certainly, more data of higher accuracy are needed in addition.

\section{Summary and outlook}\label{summary}

In this paper we have presented the electromagnetic part of a model which
is able to incorporate complete retardation in the two-body $\pi$ MEC
 and thus is 
  suited for the study of electromagnetic reactions in the two-nucleon
sector above pion threshold up to about 500 MeV excitation energy. The 
electromagnetic current of the hadronic system comprises the nucleonic 
one-body currents, the $\Delta$ excitation and meson exchange 
currents. The electromagnetic excitation of the $\Delta$ resonance has been 
fixed by fitting the $M_{1+}^{(3/2)}$ multipole of pion photoproduction. 
Within the present framework, it is possible to construct the necessary 
retarded $\pi$ MEC as an effective two-body current operator.
In this context we would like to emphasize, that we do not use any 
wave function renormalization~\cite{GaH76}, because this procedure is 
applicable in retarded models only for energies below the $\pi$ threshold. 
Consequently, 
recoil contributions besides the meson-in-flight, vertex and contact MEC 
have been taken into account in addition. The retarded $\pi$ MEC is, 
however, only partially consistent to the corresponding retarded 
$NN$ interaction. This is mainly due to the neglect of 
-- technically very complicated -- MEC of at least fourth order in the 
$\pi NN$ coupling constant which are not present in static approaches. 
On the other side, due to the use of the Siegert operators for 
the electric multipoles, such explicitly neglected mechanisms are at 
least partially taken into account. We furthermore would like to emphasize
that in  the present model {\it all} free parameters are fixed in advance 
before studying electromagnetic reactions on the two-nucleon system. In 
order to test the quality of the developed framework, we have
studied  photodisintegration of the deuteron, where until now
serious problems existed in the theoretical description above $\pi$ threshold. 
While below $\pi$ threshold, say up to about 100 MeV, both frameworks, static 
and retarded, are equivalent, this equivalence, however, breaks down 
when approaching $\pi$ threshold and even more so above, e.g., the total 
cross section in the retarded framework is considerably enhanced compared 
to the one in the static framework. Within about 10~\%, we obtain a  
satisfacory description of the experimental data over the whole energy 
region of the $\Delta$ resonance. This improvement is the combined result
of various additional independent mechanisms 
beyond the ones already considered in~\cite{WiA93}. 
These additional ingredients comprise besides the very important 
retardation effects the dissociation currents, the $\pi d$ channel and 
the explicit evaluation of the rescattering amplitude in fixing the 
$\gamma \bar{N} \Delta$ coupling. Furthermore, also the theoretical description 
of the differential cross section is considerably improved. Thus, for
the first time a model is available which is able to describe the total 
as well as the differential cross section up to about $k_{lab}$ = 450 MeV. 
However, some problems in desribing the linear photon asymmetry as well as the 
proton polarization at higher energies are still present.

\newpage
\begin{table}[btp]
\caption{Nomenclature for the pure nucleonic approaches of
 Sect.~\ref{kap6_vergleich_obepr_elster}.} 
\begin{center}
\begin{tabular}{cccc}
notation & potential & MEC & recoil current\\
\hline
N(stat,stat,0) & OBEPR & static & not included \\
N(ret,stat,0) & Elster & static & not included \\
N(ret,ret,0) & Elster & retarded & not included \\
N(ret,ret,1) & Elster & retarded & included \\
\end{tabular}
\end{center}
 \label{kap6_tab_3}
\end{table}

\begin{table}[b!]
\caption{Nomenclature for the {\it static } models of the coupled channel 
approach of Sect.~\ref{kap6_vergleich_voll}. The nomenclature for the
hadronic model follows Ref.~\protect\cite{ScA99}.
 The recoil current is always not included.
}
\begin{center}
\begin{tabular}{ccccc}
notation & hadronic model & MEC & $\gamma {\bar N} \Delta$ coupling 
& $J^{(diss)\,\, \mu}$\\
\hline
CC(stat1) & CC(stat1,$\pi$) & static & 
${\widetilde G}_{M1}^{\Delta {\bar N}}(\mbox{eff}\,2)$ & not included \\
CC(stat2) & CC(stat2,$\pi$) & static & 
${\widetilde G}_{M1}^{\Delta {\bar N}}(\mbox{eff}\,1)$ & not included \\
CC(stat3) & CC(stat,$\pi$,$\rho$,0) & static & 
${\widetilde G}_{M1}^{\Delta {\bar N}}(\mbox{eff}\,1)$ & included \\
CC(stat4) & CC(stat1,$\pi$) & static & 
${\widetilde G}_{M1}^{\Delta {\bar N}}(\mbox{eff}\,3)$ & not included \\
\end{tabular}
\end{center}
 \label{kap6_tab_stat}
\end{table}

\begin{table}[t!]
\caption{ Nomenclature for the various {\it retarded } models of the 
coupled channel approach of Sect.~\ref{kap6_vergleich_voll}. The nomenclature 
for the hadronic model follows Ref.~\protect\cite{ScA99}.
 In all cases, the $\gamma {\bar N} \Delta$ coupling 
${\widetilde G}_{M1}^{\Delta {\bar N}}(\mbox{eff}\,1)$ has been used.
 Except for CC(ret4), the retarded recoil current is always included.}
\begin{center}
\begin{tabular}{ccccc}
notation & hadronic model & MEC & remarks \\
\hline
CC(ret1) & CC(ret,$\pi$,$\rho$,-1) & retarded &  \\
CC(ret2) & CC(ret,$\pi$,$\rho$,0) & retarded &  \\
CC(ret3) & CC(ret,$\pi$,$\rho$,1) & retarded &  \\
CC(ret4) & CC(ret,$\pi$,$\rho$,0) & static & 
$J^{(\pi)/r}$, $J^{(\pi)/r}_{\Delta}$ not included \\
CC(ret5) & CC(ret,$\pi$,$\rho$,0) & retarded & $\pi d$ channel not included \\
\end{tabular}
\end{center}
 \label{kap6_tab_ret}
\end{table}

\newpage

%\newpage 

%\newpage

\end{document}